\documentstyle[12pt]{article}
\date{}

\title{String-orthogonal polynomials, String Equations and
Two-Toda Symmetries\footnote{A final version of the present paper has
appeared in:  {\em Comm. Pure and Appl. Math.}, {\bf 50} 241--290 (1997).} }

\author{M. Adler\thanks{Department of Mathematics,
Brandeis University, Waltham, Mass 02254,
USA (adler@math.brandeis.edu). The support of a National Science Foundation
grant
\# DMS 95-4-51179 is gratefully acknowledged.}~~~~~P. van
Moerbeke\thanks{Department of Mathematics, Universit\'e de Louvain, 
1348
Louvain-la-Neuve, Belgium (vanmoerbeke@geom.ucl.ac.be) and Brandeis
University, Waltham, Mass 02254, USA (vanmoerbeke@math.brandeis.edu). The
support of National  Science Foundation \# DMS 95-4-51179, Nato, FNRS and
Francqui Foundation grants is gratefully acknowledged.}}

\newcommand{\MAT}[1]{\left(\begin{array}{*#1c}}
\newcommand{\mat}{\end{array}\right)}

\newcommand{\iy}{\infty}
\newcommand{\dt}{\delta}
\newcommand{\Dt}{\Delta}
\newcommand{\al}{\alpha}
\newcommand{\be}{\beta}

\newcommand{\lb}{\lambda}
\newcommand{\sg}{\sigma}
\newcommand{\Sg}{\Sigma}
\newcommand{\Lb}{\Lambda}
\newcommand{\la}{\langle}
\newcommand{\ra}{\rangle}
\newcommand{\pp}{\ldots}
\newcommand{\pl}{\partial}

\ifx\Bbb\undefined
  \def\Bbb#1{{\bf #1}}%
  \def\BR{{\rm I\!R}}%
\else
  \newcommand{\BR}{\Bbb R}
\fi
  \newcommand{\BZ}{\Bbb Z}
  \newcommand{\BX}{\Bbb X}
  \newcommand{\BC}{\Bbb C}
  \newcommand{\BY}{\Bbb Y}

\newcommand{\rg}{\rightarrow}

\newcommand{\lrg}{\longleftrightarrow}
\newcommand{\Lrg}{\Longleftrightarrow}
\newcommand{\DR}{{\cal D}}
\newcommand{\AR}{{\cal A}}
\newcommand{\HR}{{\cal H}}
\newcommand{\LR}{{\cal L}}
\newcommand{\KR}{{\cal K}}

\newcommand{\vr}{\varepsilon}

\begin{document}
\maketitle

\vspace{0.6cm}

\noindent Table of contents:
\newline\noindent1. The 2-Toda lattice and its generic symmetries
\newline\noindent2. A larger class of symmetries for special initial conditions
\newline\noindent3. Borel decomposition of moment matrices, $\tau$-functions and
string-orthogonal polynomials
\newline\noindent4. From string-orthogonal polynomials to the
two-Toda lattice and the string equation
\newline\noindent5. Virasoro constraints on two-matrix integrals.

\vspace{1cm}

Consider a weight  $\rho (y,z) dy dz=e^{V(y,z)}dy dz$ on $\BR^2 $, with
$$
V(y,z):=V_1(y)+V_{12}
(y,z)+V_2(z):=\sum_1^{\iy}t_iy^i+\sum_{i,j\geq1} c_{ij} y^i z^j
-\sum_1^{\iy}s_iz^i,\leqno{(0.1)}
$$
the corresponding inner
product
$$\la f,g \ra = \int_{\BR^2 } dy dz \rho (y,z) f(y) g(z),\leqno{(0.2)}$$
and the moment matrix, 
$$  m_n(t,s,c) =: (\mu_{ij} )_{0 \leq ij \leq n-1} = (\la y^{i}, z^{j} \ra
)_{0 \leq i,j \leq n-1}.
\leqno{(0.3)}
$$ 
As a function of $t,s$ and $c$, the moment matrix $m_{\iy}(t,s,c)$ satisfies
the differential equations
$$\frac{\pl m_{\iy}}{\pl t_n}=\Lambda^nm_{\iy},\quad 
\frac{\pl m_{\iy}}{\pl s_n}=-m_{\iy}\Lambda^{\top^n},\quad 
\frac{\pl m_{\iy}}{\pl c_{ij}}=\Lambda^im_{\iy}\Lambda^{\top^j},
\leqno{(0.3')}
$$
where $\Lambda$ is the semi-infinite shift matrix defined below.

Consider the Borel decomposition of the semi-infinite
matrix\footnote{$\DR_{k,\ell}$ ($k<\ell\in\BZ$) denotes the set of band matrices
with zeros outside the strip
$(k,\ell).$ }
$$
m_{\iy}=S_1^{-1}S_2\mbox{  with  }S_1 \in {\cal
D}_{-\iy,0},~~S_2 \in {\cal D}_{0,\iy}
$$
with $S_1$ having $1$'s on the diagonal, and $S_2$ having $h_i$'s on the
diagonal, with (see section 3)
$$h_i=
\frac{\det m_{i+1}}{\det m_i}.$$
The Borel decomposition $m_{\iy}=S_1^{-1}S_2$ above leads to two strings
$(p^{(1)}(y),p^{(2)}(z))$ of monic polynomials in one variable, constructed, in
terms of the character 
$
\bar\chi(z)=(z^n)_{n\in\BZ, n\geq 0},
$ as follows: 
$$
p^{(1)}(y)=:S_1\bar\chi(y)  ~~~~p^{(2)}(z)=:h
(S_2^{-1})^{\top}\bar\chi(z).\leqno{(0.4)}
$$
We call these two sequences {\it string-orthogonal polynomials}; 
indeed the Borel decomposition of $m_{\iy}=S_1^{-1}S_2$ above is
equivalent to the orthogonality relations:
$$
\la p_n^{(1)},p_m^{(2)}\ra =\dt_{n,m}h_n.
$$
We show in section 4 the string-orthogonal polynomials have the following
expressions in terms of Schur differential polynomials\footnote{The Schur
polynomials $p_k$, defined by
$e^{\sum^{\infty}_1 t_i z^i }=\sum^{\infty}_0 p_k(t)z^k$ and $p_k(t)=0$ for
$k<0$, , and not to be confused with the string-orthogonal  polynomials
$p_i^{(k)}$, $k=1,2$, lead to differential polynomials
$$
p_k(\pm\tilde\pl_t)=p_k\Bigl(\pm\frac{\pl}{\pl t_1},\pm\frac{1}{2}
\frac{\pl}{\pl t_2},\pm\frac{1}{3}
\frac{\pl}{\pl t_3},...\Bigr)
$$}
$$p_n^{(1)}(y)=\sum_{0\leq k \leq n}
\frac{p_{n-k}(-\tilde\pl_t)\det m_n(t,s,c)}{\det m_n(t,s,c)}y^k \mbox{~,~}
p_n^{(2)}(z)=\sum_{0\leq k \leq n}
\frac{p_{n-k}(\tilde\pl_s)\det m_n(t,s,c)}{\det m_n(t,s,c)}z^k
$$
To the best of our
knowledge, string-orthogonal polynomials were considered for the first time, in
the context of symmetric weights
$\rho(y,z)dydz$, by Mehta (see [M] and [CMM]). 

The {\em main message} of this work is to show (i) that the expressions 
$\det m_N$ satisfy the KP-hierarchy in $t$ and $s$ for each $N=1,2,...$,
(section 3) (ii) that the $\det m_N$ hang together in a very specific way
(they form the building blocks of the 2-Toda lattice) (section 4), (iii) that
$\det m_N$ satisfies an additional Virasoro-like algebra of partial
differential equations (section 5), as a result of so-called string equations
(section 4). In [AvM2], we have obtained similar results for moment matrices
associated with general weights on $\Bbb R$ (thus connecting with the
standard theory of orthogonal polynomials), rather than on
$\Bbb R^2$. It would not be difficult to generalize the results of this paper
to more general weights
$\rho(y,z)dy\,dz$ on $\Bbb R^2$, besides those of (0.1).

In terms of the matrix operators
$$\Lambda :=
\left(\begin{tabular}{llll}
0&1 & & \\
 0&0&1& \\
 &0 &0& \\
 & & &$\ddots$ 
\end{tabular}
\right)\mbox{  and  }\vr:=\left(\begin{tabular}{llll}
0&0 & & \\
1&0&0 & \\
 &\hfill 2&0&\\
 & &   &$\ddots$ 
\end{tabular}
\right)
$$
acting on $\bar \chi$ as
$$
\Lambda\bar\chi(z)=z\bar\chi(z), ~\vr\bar\chi(z)=\frac{\pl}{\pl z}\bar\chi(z), 
$$
the matrices
$$
L_1:=S_1\Lb S_1^{-1}, L_2:=S_2\Lb^{\top} S_2^{-1}, Q_1:=S_1 \vr S_1^{-1},  
Q_2:=S_2 \vr^{\top} S_2^{-1}
$$ 
interact with the vector of string-orthogonal
polynomials, as follows:
$$
L_1p^{(1)}(y)=yp^{(1)}(y)~~~~~~Q_1p^{(1)}(y)=\frac{d}{dy}p^{(1)}(y)\leqno{(0.5)}
$$
$$
~~~hL_2^{\top}h^{-1}p^{(2)}(z)=zp^{(2)}(z)~~~~~~ 
hQ_2^{\top}h^{-1}p^{(2)}(z)=\frac{d}{dz}p^{(2)}(z).
$$
 The semi-infinite matrix $L_1$ (respectively $L_2$) is lower-triangular (resp.
lower-triangular), with one subdiagonal above (resp. below); $Q_1$ (resp.
$Q_2$) is strictly lower-triangular (resp. strictly upper-triangular). In Theorem
4.1 we prove the matrices
$L_i$ and
$Q_i$ satisfy so-called {\em ``string equations"}
$$
Q_1+\frac{\pl V}{\pl y}(L_1,L_2)=0,\quad\quad \quad Q_2+
\frac{\pl V}{\pl z}( L_1,
L_2)=0.\leqno(0.6)
$$
When
$$
V(y,z)=\sum_1^{\ell_1}t_iy^i+cyz
-\sum_1^{\ell_2}s_iz^i,
$$
then (0.6) implies that $L_1$ is a $\ell_2+1$-band matrix, and $L_2$ is a
$\ell_1+1$-band matrix.

Moreover, as a function of $(t,s)$, the couple $L:=(L_1,L_2)$ satisfies the
{\em `` two-Toda lattice equations"}, and as a function of $c_{\al,\be}$, $L$
satisfies another hierarchy of commuting vector fields; so, in terms of an
appropriate Lie algebra splitting $(~)_+$ and $(~)_-$, to be explained in
section 1, we have
$$
\frac{\pl L}{\pl t_n}=[(L^n_1,0)_+,L]\quad \frac{\pl L}{\pl
s_n}=[(0,L_2^n)_+,L],\quad \frac{\pl L}{\pl
c_{\al,\be}}=-[(L_1^{\al}L_2^{\be},0)_-,L],\leqno(0.7)
$$
and, what is equivalent, the moment matrix $m_{\iy}$ satisfies the
differential equations (0.3'), 
with solution (thinking of $t_n=c_{n0}$ and $s_n=-c_{0n}$):
$$
m_{\iy}(t,s,c)=\renewcommand{\arraystretch}{0.5}
\begin{array}[t]{c}
\sum\\
{\scriptstyle (r_{\al\be})_{\al,\be\geq 0}\in\BZ^{\iy}}\\
{\scriptstyle \,\,\,\,\,_{(\al,\be)\neq (0,0)}}
\end{array}
\renewcommand{\arraystretch}{1}
\left(\prod_{(\al,\be)}\frac{c_{\al\be}^{r_{\al\be}}}{r_{\al\be}!}\right)
\Lambda^{\sum_{\al\geq 1}\al
r_{\al\be}}m_{\iy}(0,0,0)\Lambda^{\top\sum_{\be\geq 1}\be
r_{\al\be}};
$$
in particular
$$
m_{\iy}(t,s,0)=e^{\sum_1^{\iy} t_n
\Lambda^n}m_{\iy}(0,0,0)e^{-\sum_1^{\iy} s_n
\Lambda^{\top n}} .
$$
Thus the integrable system under
consideration in this paper is a $c_{\al\be}$-deformation of the
2-Toda lattice, which itself is an isospectral deformation of a
couple of matrices
$(L_1,L_2)$, in general bi-infinite, depending on two sequences of times
$t_1,t_2,\ldots$ and
$s_1, s_2,\ldots$.

Moreover, the determinant of the moment matrix has many different
expressions; in particular, in terms of the moment
matrix at $t=s=0$, using the matrix $E_N(t):=$ (the first $N$ rows of
$e^{\sum_1^{\iy}t_n\Lb^n}$) of Schur polynomials
$p_n(t)$ (see section 3). It can also be expressed 
as a 2-matrix integral reminiscent of
2-matrix integrals in string theory and in terms of the diagonal elements
$h_i$ of the upper-triangular matrix $S_2$:

\medbreak
\noindent (0.8)
\begin{eqnarray*}
N!\det m_N(t,s,c) 
&=&N!\det\Bigl(E_N(t)m_{\iy}(0,0,c)E_N(-s)^{\top}\Bigr)\\
&=&\int\int_{u,v\in\BR^N}e^{\sum^N_{k=1}V(u_k,v_k)}\prod_{i<j}
(u_i-u_j)\prod_{i<j}(v_i-v_j)dudv \\
&=&\prod_0^{N-1}h_i(t,s,c)\\ &=:&\tau_N(t,s,c).
 \end{eqnarray*}
Finally $\tau_N(t,s,c)$ is a
 {\em ``$\tau$-function"} in the sense of Sato, separately in
$t$ and $s$\footnote{Sato's $\tau$-function $\tau(t)$ in $t \in \BC^{\iy}$
is the determinant of the projection $e^{-\sum t_i z^i} W \rightarrow
H_+=\{1,z,z^2,\ldots\}$, where $W$ is a fixed span of functions in $z$ with
poles at $z=\iy$ of order $k=0,1,2,
\ldots$. Equivalently, a $\tau$-function satisfies the bilinear
relations  
$$
\oint_C\tau(t-[z^{-1}])\tau(t'+[z^{-1}])e^{\sum_1^{\iy}
(t_i-t'_i)z^i}dz=0,
$$ where $C$ is a small contour about $z=\iy$ and where
$[\al]:=\Bigl(\al,\frac{\al^2}{2},\frac{\al^3}{3},...\Bigr)\in\Bbb C^{\iy}$. The
bilinear relations imply that
$\tau(t)$ satisfies the KP-hierarchy.}. 

The string equations (0.6) play an important role: they have many consequences!
In Theorem 5.1, we show
$\tau_N=\det m_N(t,s,c)$ satisfies the following set of constraints\footnote{in
terms of the customary Virasoro generators in $t_1,t_2,...$:
\begin{eqnarray*}
J_n^{(0)}&=&\dt_{n0},\hspace{2cm}J_n^{(1)}=\frac{\pl}{\pl
t_n}+(-n)t_{-n},\hspace{2cm} J_0^{(1)}=0\\
J_n^{(2)}&=&\sum_{i+j=n}:J_i^{(1)}J_j^{(1)}:=\sum_{i+j=n}\frac{\pl^2}{\pl
t_i\pl t_j}+2\sum_{-i+j=n}it_i\frac{\pl}{\pl t_j}+\sum_{-i-j=n}(it_i)(jt_j),
\end{eqnarray*}
where ``:\,\,:" denotes normal ordering, i.e., always pull differentiation to
the right, and where a symbol $=0$, whenever it does not make sense; e.g.,
$\pl/\pl t_n=0$ for $n\leq 0$. We also define Virasoro generators in $s$,
namely
$\tilde J_n^{(k)}:=J_n^{(k)}\Bigl|_{t\rg s}$.}
$$
\Biggl(J_i^{(2)}+(2 N+i+1)J_i^{(1)}
+N(N+1) J_i^{(0)}+2\sum_{r,s\geq
1}rc_{rs}\frac{\pl}{\pl
c_{i+r,s}}\Biggr)\tau_N=0
\leqno{(0.9)}
$$
$$
\Biggl(\tilde
J_i^{(2)}-(2 N+i+1)\tilde J_i^{(1)}
+N(N+1) \tilde J_i^{(0)}+2\sum_{r,s\geq
1}sc_{rs}\frac{\pl}{\pl c_{r,s+i}}\Biggr)\tau_N=0,
$$ \hfill for $i\geq -1$ and $N\geq 0.$

\noindent When $V_{12}=cyz$, the relations (0.9) reduce to an inductive system
of partial differential equations in $t$ and $s$, for $\tau_0,\tau_1,
\tau_2, \ldots$, (Corollary 5.1.1)
$$\leqno{(0.10)}$$
\begin{eqnarray*}
\left(J_i^{(2)}+(2N+i+1)J_i^{(1)}+N(N+1)J_i^{(0)}\right)\tau_N+
2c~p_{i+N}
(\tilde\pl_t)p_{N}(-\tilde\pl_s)\tau_1\circ\tau_{N-1}&=&0  \\
\\
\left(\tilde J_i^{(2)}-(2N+i+1)\tilde
J_i^{(1)}+N(N+1)\tilde J_i^{(0)}\right)\tau_N +2c~p_{N}
(\tilde\pl_t)p_{i+N}(-\tilde\pl_s)\tau_1\circ\tau_{N-1}&=&0
\end{eqnarray*} 
\hfill for $i\geq -1$ and $N\geq 1.$ 

\noindent involving the Hirota operation\footnote{Given two
differential polynomials
$p(\pl_t)$ and
$q(\pl_s)$, and functions $f(t_1,t_2,...;s_1,s_2,...)$,
$g(t_1,t_2,...;s_1,s_2,...)$, define the Hirota operation for shifts
$y=(y_1,y_2,...)$ and
$z=(z_1,z_2,...)$:
$$
p(\pl_t)q(\pl_s)f\circ g(t,s):=p(\frac{\pl}{\pl y})q(\frac{\pl}{\pl
z})f(t+y,s+z)g(t-y,s-z)\Bigl|_{y=z=0}.
$$}; to be precise, the expression $p_{n}
(\tilde\pl_t)p_{m}(-\tilde\pl_s)f\circ g$, for $\tilde\pl_t$ and
$\tilde\pl_s$ spelled out in footnote 2, is defined as the coefficient in the 
$(u,v)$-Taylor expansion of 
\begin{eqnarray*}
f(t+[u],s-[v]) g(t-[u],s+[v])&=&\sum_{n,m \geq 0} u^n v^m p_n(\tilde\pl_t)
p_m(-\tilde\pl_s)f \circ g  \\
&=& \sum_{n,m \geq 0} u^n v^m  
\renewcommand{\arraystretch}{0.5}
\begin{array}[t]{c}
\sum\\
{\scriptstyle i+i'= n}\\
{\scriptstyle j+j'=m}\\
{\scriptstyle i,i',j,j'\geq 0}
\end{array}
\renewcommand{\arraystretch}{1}p_{i}
(\tilde\pl_t)p_{j}(-\tilde\pl_s)f~. 
p_{i'} (-\tilde\pl_t)p_{j'}(\tilde\pl_s)g.
\end{eqnarray*}

\noindent Also, in terms of W-generators, to be defined in (1.36),
the matrix integral (0.8) satisfies (Theorem 5.2):
$$
\Bigl(\sum_{k=0}^{i\wedge
j}\bar\al_k^{(i,j)}W_{N,i-j}^{(j-k+1)}+\sum_0^i\bar\be_k^{(i)}
\tilde W_{N-1,j-i}^{(i-k+1)}
\Bigr)\tau_N=\frac{i!}{(-c)^i}\dt_{ij}\tau_N,\quad\quad
i,j\geq 0 ,\leqno{(0.11)}$$ 
where $\bar\al_k^{(i,j)}$ and $\bar\be_k^{(i)}$ are numbers defined in (5.10). A
relation of this type was conjectured by Morozov [M].

The technique employed here is the one of symmetry vector
fields of integrable systems  (KP, Toda, two-Toda, etc...), as developped in
[ASV1] and [AV2]; they are non-local and time-dependent vector fields
transversal to and commuting with the integrable hierarchy, when
acting on the ({\it explicitly}) time-dependent wave functions;
however bracketing a symmetry
with a Toda vector field (0.7), which acts on the ({\it
implicitly}) time-dependent $L$, yields another vector field in the
hierarchy.  In [ASV], we have shown that for generic initial conditions
the 2-Toda lattice admits a
$w_{\infty}\times w_{\infty}$ algebra of symmetries, i.e., an algebra without central
extension. 

In this study, we consider special initial conditions, 
preserved by the Toda flows;
we show they admit a much wider algebra of symmetries, to be described in Theorem
2.1. This big algebra contains not only the algebra $w_{\infty}\times w_{\infty}$
above, but also a Kac-Moody extension $w_{\iy}\otimes\BC(h,h^{-1})$ of $w_{\infty}$
(in two different ways). Via the Adler-Shiota-van Moerbeke formula, symmetries at the
level of the wave functions induce symmetries at the level of the
$\tau$-function; these symmetries form an algebra with central extension. 
van de Leur [vdL1,2] has a striking representation-theoretical generalization of
the ASV-formula; it is an open question to understand whether it includes the
wider class of symmetries under consideration here.

The string equations (0.6) are equivalent to the vanishing of a whole algebra of
symmetries, viewed as vector fields on the manifold of wave functions; they lead,
upon using the ASV-formula, to the algebra of constraints (0.9) and (0.10) (on
the 2-matrix integral), viewed as vector fields on the manifold of 
$\tau$-functions, and this in
a straightforward, conceptual and precise fashion.

We wish to thank S. D'Addato for a superb typing job.

\section{The 2-Toda lattice and its generic symmetries}

Define the column vector $\chi(z)=(z^n)_{n\in\BZ}$, and matrix operators $\Lambda$, 
$\Lambda^*$, $\vr,\vr^*$ defined as follows:
$$\Lambda \chi(z)=z\chi(z), ~\vr\chi(z)=\frac{\pl}{\pl z}\chi(z), 
$$
$$\Lambda^* \chi(z)=z^{-1} \chi(z), ~\vr^*\chi(z)=\frac{\pl}{\pl z^{-1}}\chi(z).
$$
Note that
$$
\Lambda^*=\Lambda^{\top}=\Lambda^{-1},~\vr^*=-\vr^{\top}+\Lambda,
$$
and
$$
\left\{
\begin{array}{l}
\Lambda^{\top}\chi(z^{-1})=z\chi(z^{-1}),~\Lb\chi
(z^{-1})=z^{-1}\chi(z^{-1}),\\
\vr^{\top}\chi(z^{-1})=z^{-1}\chi(z^{-1})-
\frac{\pl}{\pl z}\chi(z^{-1})\\
\vr^{*\top}\chi(z^{-1})=z\chi(z^{-1})-
\frac{\pl}{\pl z^{-1}}\chi(z^{-1}).
\end{array}
\right.\leqno(1.1)
$$

Consider the splitting
of the algebra ${\cal D}$ of pairs $(P_1,P_2)$ of
infinite ($\BZ \times \BZ$) matrices such that $(P_1)_{ij}=0$ for
$j-i\gg0$ and $(P_2)_{ij}=0$ for $i-j\gg0$, used in [ASV2]; to wit:
\begin{eqnarray*}
\DR&=&\DR_++\DR_-,\\
\DR_+&=&\bigl\{(P,P)\bigm|P_{ij}=0\hbox{ if }|i-j|\gg0\bigr\}
=\bigl\{(P_1,P_2)\in\DR\bigm| P_1=P_2 \bigr\},\\
\DR_-&=&\bigl\{(P_1,P_2)\bigm|(P_1)_{ij}=0\hbox{ if }j\ge i,\
(P_2)_{ij}=0\hbox{ if }i>j\bigr\},
\end{eqnarray*}
with $(P_1,P_2)=(P_1,P_2)_++(P_1,P_2)_-$ given by
$$
\begin{array}{c}
(P_1,P_2)_+=(P_{1u}+P_{2\ell},P_{1u}+P_{2\ell}),
\\[3mm]
(P_1,P_2)_-=(P_{1\ell}-P_{2\ell},P_{2u}-P_{1u});
\end{array}
\leqno(1.2)
$$
$P_u$ and $P_\ell$ denote the upper (including diagonal)
and strictly lower triangular parts of the matrix $P$, respectively.

The two-dimensional Toda lattice equations
$$
{\pl L\over\pl t_n}=\bigl[\bigl(L_1^n,0\bigr)_+,L\bigr]\quad\hbox{and}\quad
{\pl L\over\pl s_n}=\bigl[\bigl(0,L_2^n\bigr)_+,L\bigr]\quad n=1,2,\dots
\leqno(1.3)
$$
are deformations of a pair of infinite matrices
$$
L=(L_1,L_2)
=\Bigl(\sum_{-\iy<i\le1}a^{(1)}_i\Lb^i,\sum_{-1\le i<\iy}a^{(2)}_i\Lb^i\Bigr)
\in\DR,
\leqno(1.4)
$$
where $\Lb=(\dt_{j-i,1})_{i,j\in\BZ}$, and
$a_i^{(1)}$ and $a_i^{(2)}$ are
diagonal matrices depending on $t=(t_1,t_2,\dots)$ and $s=(s_1,s_2,\dots)$,
such that
$$
a_1^{(1)}=I\quad\hbox{and}\quad \bigl(a_{-1}^{(2)}\bigr)_{nn}\ne
0\quad\mbox{for all}\,\, n. $$

In analogy with Sato's theory, Ueno and Takasaki [U-T] show a
solution $L$ of (1.3) has the representation
$$
L_1=W_1\Lambda W_1^{-1}=S_1\Lambda S_1^{-1},~~L_2=W_2
\Lambda^{-1} W_2^{-1}=S_2\Lambda^{-1} S_2^{-1}
$$
in terms of two
pairs of wave operators
$$
\left\{
\begin{array}{l}
S_1=\sum_{i\leq 0}c_i(t,s)\Lambda^i,\quad S_2=\sum_{i\geq 0}
c'_i(t,s)\Lambda^i\\
c_i,c'_i:\mbox{\,diagonal matrices}, c_0=I,(c'_0)_{ii}
\neq 0,\mbox{\, for all}\,\, i
\end{array}
\right.
$$
and
$$
W_1=S_1(t,s)e^{\sum_1^{\iy}t_k\Lambda^k},\quad W_2=S_2(t,s)
e^{\sum_1^{\iy}s_k\Lambda^{-k}}.\leqno(1.5)
$$

One also introduces a pair of
wave vectors
$\Psi=(\Psi_1,\Psi_2)$, and a pair of adjoint wave
vectors, $\Psi^*=(\Psi^*_1,\Psi^*_2)$, instead of a single wave function and a
single adjoint wave function:
$$
\left\{
\begin{array}{l}
\Psi_1(t,s,z)=W_1\chi(z)=e^{\sum_1^{\iy}t_kz^k}S_1\chi(z)\\
\Psi_2(t,s,z)=W_2\chi(z)=e^{\sum_1^{\iy}s_kz^{-k}}S_2\chi(z)
\end{array}
\right.\leqno{(1.6)}
$$
and
$$
\left\{
\begin{array}{l}
\Psi_1^*=(W_1^{\top})^{-1}\chi^*(z)=
e^{-\sum_1^{\iy}t_kz^k}(S_1^{\top})^{-1}\chi(z^{-1})\\
\Psi_2^*=(W_2^{\top})^{-1}\chi^*(z)=
e^{-\sum_1^{\iy}s_kz^{-k}}(S_2^{\top})^{-1}\chi(z^{-1}).
\end{array}
\right.\leqno(1.7)
$$
The wave functions $\Psi$ and $\Psi^*$ evolve in $t$ and $s$ 
according to the following differential equations\footnote{Here the action 
is viewed componentwise, e.g., $(A,B)\Psi=(A\Psi_1,B\Psi_2)$ or
$(z,z^{-1})\Psi=(z\Psi_1,z^{-1}\Psi_2)$.}:
$$
\left\{
\begin{array}{l}
\frac{\pl \Psi}{\pl t_n}=(L_1^n,0)_+ \Psi=((L_1^n)_u,(L_1^n)_u)\Psi\\
\frac{\pl \Psi}{\pl s_n}=(0,L_2^n)_+ \Psi=((L_2^n)_{\ell},(L_2^n)_{\ell})\Psi
 \end{array}
\right.\leqno(1.8)
$$
$$
\left\{
\begin{array}{l}
\frac{\pl}{\pl t_n}\Psi^* =((L_1^n,0)_+)^{\top} \Psi^{\ast}\\
\frac{\pl}{\pl s_n}\Psi^* =-((0,L_2^n)_+)^{\top}\Psi^*.
\end{array}
\right.\leqno(1.9)
$$
Besides $L=(L_1,L_2)=(S_1\Lb S_1^{-1},S_2\Lb^{\top}S_2^{-1})$, the
operators 
$$
L^*=(L^*_1,L_2^*)=(L_1^{\top},L_2^{\top})
$$
$$M=(M_1,M_2):=(W_1\vr
W_1^{-1}, W_2 \vr^{\ast}
W_2^{-1})
=\Bigl(S_1(\varepsilon+\sum_1^{\iy}kt_k\Lb^{k-1}\Bigr)S_1^{-1},
S_2\Bigl(\varepsilon^*+\sum_1^{\iy}ks_k\Lb^{\top
k-1})S_2^{-1}\Bigr)
$$
$$
M^*=(M_1^*,M_2^*)=(-M_1^{\top}+L_1^{\top
-1},-M_2^{\top}+L_2^{\top
-1})
$$
satisfy, in view of (1.1):
$$
\begin{array}{c}
        L\Psi=(z,z^{-1})\Psi,\quad
        M\Psi=\Bigl({\pl\over\pl z},{\pl\over\pl(z^{-1})}\Bigr)\Psi,\quad
        [L,M]=(1,1),
        \\
 L^*\Psi^*=(z,z^{-1})\Psi^*,\quad
        M^*\Psi=\Bigl({\pl\over\pl z},
							{\pl\over\pl(z^{-1})}\Bigr)\Psi^*,\quad
        [L^*,M^*]=(1,1).
        \\[5mm]
       
\end{array}
\leqno(1.10)
$$
The operators $L,M,L^{\top},M^{\top}$ and $W:=(W_1,W_2)$ evolve according to
$$
\left\{
\begin{array}{l}
\frac{\pl}{\pl t_n}\MAT{1}L\\M\mat=\Biggl[(L_1^n,0)_+,\MAT{1}L\\M\mat\Biggr]\\
\\
\frac{\pl}{\pl s_n}\MAT{1}L\\M\mat=\Biggl[(0,L_2^n)_+,\MAT{1}L\\M\mat\Biggr]
 \end{array}
\right.\leqno(1.11)
$$

$$
\left\{
\begin{array}{l}
\frac{\pl}{\pl t_n}\MAT{1}L^{\top}\\M^{\top}\mat=\Biggl[-\Bigl(
(L_1^n,0)_+\Bigr)^{\top},\MAT{1}L^{\top}\\M^{\top}\mat\Biggr]\\
\\
\frac{\pl}{\pl
s_n}\MAT{1}L^{\top}\\M^{\top}\mat=\Biggl[-\Bigl((0,L_2^n)_+\Bigr)
^{\top},\MAT{1}L^{\top}\\M^{\top}\mat\Biggr]
\end{array}
\right.\leqno(1.12)
$$
$$
\left\{
\begin{array}{l}
\frac{\pl W}{\pl t_n}=(L_1^n,0)_+ W,\\
\\
\frac{\pl W}{\pl s_n}=(0,L_2^n)_+ W
 \end{array}
\right.\leqno(1.13)
$$
and consequently
\begin{eqnarray*}
\frac{\pl W_1^{-1}W_2}{\pl
t_n}&=&-W_1^{-1}(L_1^n)_uW_1W_1^{-1}W_2+W_1^{-1}(L_1^n)_uW_2=0\\
\frac{\pl W_1^{-1}W_2}{\pl s_n}&=&0.
\end{eqnarray*}
This implies that
$$
W_1^{-1}W_2(t,s)=W_1^{-1}W_2(0,0)\leqno(1.14)
$$
and thus
$$(S_1^{-1}S_2)(t,s)=e^{\Sg t_n\Lb^n}(S_1^{-1}S_2)(0,0)e^{-\Sg s_n(\Lb^{\top})^n}.
\leqno(1.15)
$$

Ueno and Takasaki [U-T] show that the 2-Toda deformations of
$\Psi$, and hence $L$, can ultimately all be expressed in terms of one sequence of 
$\tau$-functions $$\tau(n,t,s)=
\tau_n(t_1,t_2,\dots;s_1,s_2,\dots)=
\det[(S_1^{-1}S_2(t,s))_{i,j}]_{-\infty \leq i,j\leq n-1},\quad n\in\BZ:
$$
to wit:
$$
\Psi_1(t,s,z)=\biggl(
        {e^{-\eta}\tau_n(t,s)\over\tau_n(t,s)}e^{\sum t_iz^i}z^n
\biggr)_{n\in\BZ},\quad
\Psi_2(t,s,z)=\biggl(
        {e^{-\tilde\eta}\tau_{n+1}(t,s)\over\tau_n(t,s)}e^{\sum s_iz^{-i}}z^n
\biggr)_{n\in\BZ}
\leqno(1.16)
$$
$$
\Psi^*_1(t,s,z)=\Biggl(\frac{e^{\eta}\tau_{n+1}(t,s)}
{\tau_{n+1}(t,s)}e^{-\Sg t_iz^i}z^{-n}\Biggr)_{n\in\BZ},\quad 
\Psi^*_2(t,s,z)=\Biggl(\frac{e^{\tilde\eta}\tau_n(t,s)}
{\tau_{n+1}(t,s)}e^{-\Sg s_iz^{-i}}z^{-n}\Biggr)_{n\in\BZ}.
\leqno(1.17)
$$
where
$$
        \eta=\sum_1^\iy{z^{-i}\over i}{\pl\over\pl t_i}
        \quad\hbox{and}\quad
        \tilde\eta=\sum_1^\iy{z^i\over i}{\pl\over\pl s_i},
\leqno(1.18)
$$
so that
$$
        e^{a\eta+b\tilde\eta}f(t,s)=f(t+a[z^{-1}],s+b[z])
\leqno(1.19)
$$
with $[\al]:=(\al,\al^2/2,\al^3/3,\dots)$. Here the labeling of $\Psi^*_i$ is
slightly different from the one of [U-T].

The symmetries of the 2-Toda hierarchy are conveniently expressed
in terms of the operators $L$ and $M$.
In view of the relation
$$
z^\al\Bigl({\pl\over\pl z}\Bigr)^\be\Psi_1
=M_1^\be L_1^\al\Psi_1,\quad
u^\al\Bigl({\pl\over\pl u}\Bigr)^\be\Bigr|_{u=z^{-1}}\Psi_2
=M_2^\be L_2^\al\Psi_2,
\leqno(1.20)
$$
the Lie algebra
$$
w_{\iy}:=\mbox{span}\Bigl\{
        z^{\al}\Bigl({\pl\over\pl
z}\Bigr)^{\be}\Bigm|\al,\be\in\BZ,\be\ge 0 \Bigr\}
$$
comes naturally into play. To be precise, denoting by $\phi$ the algebra
antihomomorphism
$$
\phi\colon w_{\iy}\times w_\iy \to\DR:
\left\{
\begin{array}{l}
        \bigl(z^\al(\pl/\pl z)^\be,0\bigr)
        \mapsto\bigl(M_1^\be L_1^\al,0\bigr)\\
	\\
        \bigl(0,u^\al(\pl/\pl u)^\be\bigr)
        \mapsto\bigl(0,M_2^\be L_2^\al\bigr)
\end{array}
\right.
$$
we have a Lie algebra antihomomorphism\footnote{Thus $-\BY$ is a Lie
algebra homomorphism; note we shall more often denote $\BY_p$ by $\BY_{\phi(p)}$.}
\begin{eqnarray*}
\BY\colon w_\iy\times w_\iy&\to&\bigl\{\hbox{symmetries on
$\Psi,\Psi^*,L,L^{\top},M,M^{\top}$}\bigr\}\\
p&\longmapsto&\left\{
        \protect{
        \begin{array}{l}
                \BY_p\Psi
                =-\phi(p)_-\Psi,~\BY_p\Psi^*
                =\left(\phi(p)_-\right)^{\top}\Psi^*,\\
                \BY_pL
                =\bigl[-\phi(p)_-,L\bigr],~\BY_pL^{\top}
                =\bigl[\left(\phi(p)_-\right)^{\top},L\bigr],\\
                \BY_pM
                =\bigl[-\phi(p)_-,M\bigr],~\BY_pM^{\top}
                =\bigl[\left(\phi(p)_-\right)^{\top},M^{\top}\bigr].
        \end{array}
        }
\right.
\end{eqnarray*}

\noindent So for any admissible\footnote {The products $M_i^{\beta} L_i^{\alpha}$
are admissible, but, in this generality, the product $L_1 L_2$ obviously does
not make sense.} polynomial $q$ of $L$ and $M$, we have $$
  \BY_p(q)=[-p_-,q]
  \quad\mbox{and}\quad 
  \BY_p \Psi=-p_-\Psi.\leqno(1.21)$$
To prove
$$
[\BY_{p_1},\BY_{p_2}]=-\BY_{[p_1,p_2]},
$$
one computes
$$
[\BY_{p_1},\BY_{p_2}]\Psi=-Z_1\Psi,
$$
where, upon using $\DR_-$ and $\DR_+$ are Lie subalgebras,
\begin{eqnarray*}
Z_1&:=&-\Bigl(\BY_{p_1}(p_2)\Bigr)_-+\Bigl(\BY_{p_2}(p_1)\Bigr)_-
-[p_{1-},p_{2-}]\\
&=&[p_{1-},p_2]_-+[-p_{2-},p_1]_--[p_{1-},p_{2-}]=[p_1,p_2]_-.
\end{eqnarray*}
Moreover
$$ [\BY_p,\frac{\pl}{\pl t_n}]\Psi=[\BY_p,\frac{\pl}{\pl s_n}]\Psi =0;
$$
indeed, from (1.8), the vector field,
$$
\BZ_q\Psi=q_+\Psi,\quad\BZ_qp=[q_+,p]
\leqno(1.22)
$$
represents $\frac{\pl}{\pl t_n}$ for $q=(L_1^n, 0)$ and $\frac{\pl}{\pl s_n}$ for
$q=(0,L_2^n)$; now one checks
 $$
[\BY_p,\BZ_q]\Psi=-Z_2\Psi\leqno(1.23)
$$
where
$$
Z_2=(\BZ_q(p))_-+(\BY_p(q))_+-[q_+,p_-]=[q_+,p]_-
+[-p_-,q]_+-[q_+,p_-]=0.
$$

\medbreak\noindent \underline{Remark 1}: Some explanation is necessary to 
understand the equation above $\BY_p\Psi^*
=\left(\phi(p)_-\right)^{\top}\Psi^*$: the flows on the
transposed matrices $L^{\top}, M^{\top}$ are immediately obtained
from the original flows by just taking transposes; the flows on the
adjoint wave operator $\Psi^*$ are obtained as follows: since
$$
\left\{
\begin{array}{l}
(\Psi_1,\Psi_2)=(W_1,W_2)\chi(z)\\
(\Psi_1^*,\Psi_2^*)=((W_1^{-1})^{\top},(W_2^{-1})^{\top})\chi(z^{-1}),
\end{array}
\right.\leqno(1.24)
$$
we have for any deformation $^{\prime}$, that
$$
\Psi'_i=W'_i\chi(z),\quad\Psi^{*'}_i=(W_i^{-1})^{\top '}\chi(z).
$$
Thus, we have the straightforward equivalences:
$$
W'_i=A_iW_i\Lrg\Psi'_i=A\Psi_i
$$
and 
$$(W_i^{-1})^{\top'}=-A^{\top}(W_i^{-1})^{\top}\Lrg\Psi_i^{*'}=-A^{\top}
\Psi^*_i;
$$
so, we conclude
$$
\Psi'_i=A\Psi_i\Lrg\Psi_i^{*'}=-A^{\top}\Psi_i^*.\leqno(1.25)
$$

\medbreak\noindent \underline{Remark 2}: Spelling out the symmetries (1.21) on $L$
and $\Psi$, one finds $$
\BY_{M_i^{\al}L_i^{\be}}L=(-1)^{i-1}\left[\left(-(M_i^{\al}L_i^{\be})_{\ell},
(M_i^{\al}L_i^{\be})_u \right),L\right]\leqno(1.26)
$$
$$
\BY_{M_i^{\al}L_i^{\be}}\Psi=(-1)^{i-1}
\left(-(M_i^{\al}L_i^{\be})_{\ell},
(M_i^{\al}L_i^{\be})_u\right)\Psi,
\leqno(1.27)
$$
and the equivalent equations, 
upon taking the transpose of the previous ones,
$$
\BY_{M_i^{\al}L_i^{\be}}L^{\top}=(-1)^{i-1}
\left[\left((M_i^{\al}L_i^{\be})_{\ell},
-(M_i^{\al}L_i^{\be})_u \right)^{\top},L^{\top}\right]$$
$$\leqno(1.27')$$
$$
\BY_{M_i^{\al}L_i^{\be}}\Psi^*=
(-1)^{i-1}\left((M_i^{\al}L_i^{\be})_{\ell},
-(M_i^{\al}L_i^{\be})_u\right)^{\top}\Psi^{\ast}.
$$
\bigbreak

 The action of a vector
field on $\Psi$, commuting with the Toda flows,
 induces an action on $\tau$ via (1.16), thus leading to
the following relation between logarithmic derivatives of $\Psi$ and
$\tau$:
$$
\left(\frac{\Psi'_1}{\Psi_1}\right)_n=(e^{-\eta}-1)\frac{\tau'_n}{\tau_n},\quad
\left(\frac{\Psi'_2}{\Psi_2}\right)_n=(e^{-\tilde\eta}-1)
\frac{\tau'_{n+1}}{\tau_{n+1}}+\Bigl(
\frac{\tau'_{n+1}}{\tau_{n+1}}
-\frac{\tau'_n}{\tau_n}\Bigr).
\leqno{(1.28)}
$$

\proclaim Proposition 1.1. Given the symmetry vector field $\BY_p$, the following
relation holds:
 $$
\BY_p\log \frac{\tau_{n+1}}{\tau_n}=p_{nn}\leqno(1.29)
$$

\noindent\underline{Proof}: On the one hand, taking into account
the expression (1.16) for $\Psi_2$, we have, in view of the fact
that $e^{-\tilde \eta} f(t,s)=f(t,s-[z])$, the following Taylor
expansion about $z=0$:
$$
\begin{array}{ll}
\Bigl(\frac{\BY_p\Psi_2}{\Psi_2}\Bigr)_n&=\Bigl(\frac{p_u\Psi_2}{\Psi_2}
\Bigr)_n\\
\\[3mm]
&=\frac{(p_{nn}
+O(z))(\frac{\tau_{n+1}(t,s)}
{\tau_n(t,s)}+O(z))z^n}{(\frac{\tau_{n+1}(t,s)}
{\tau_n(t,s)}+O(z))z^n}\\
\\[3mm]
&=p_{nn}+O(z).
\end{array}\leqno{(\rm a)}
$$
On the other hand,
$$
(e^{-\tilde \eta}-1)\frac{\tau'_{n+1}}{\tau_{n+1}} +
(\frac{\tau'_{n+1}}{\tau_{n+1}}-\frac{\tau'_n}{\tau_n}) =
\frac{\tau'_{n+1}}{\tau_{n+1}}  -\frac{\tau'_n}{\tau_n} +
O(z)=(\log\frac{\tau_{n+1}}{\tau_n})'+O(z).
\leqno{\rm (b)}$$
According to (1.28), the expressions (a) and (b) are equal, leading to (1.29).

\bigbreak

In order to spell out the precise relationship between the symmetry 
vector fields on $\Psi$, and those acting on $\tau$, we need to 
discuss generating functions for the symmetries.

A generating function of the {\bf symmetries on the $\Psi$-manifold} is given by
$$
\BY_N\Psi=-N_-\Psi\leqno(1.30)
$$
which naturally leads to the symmetry on the $L$-manifold
expressed in the Lax form
$$
\BY_NL=[-N_-,L],\leqno(1.31)
$$
with
$$
N=(N_1,0)\ \ \hbox{or}\ \ (0,N_2),\leqno(1.32)
$$
$$
N_i:=(\mu-\lb)e^{(\mu-\lb)M_i}\dt(\lb,L_i)=\sum_{k=1}^\iy{(\mu-\lb)^k\over k!}
\sum_{\ell=-\iy}^\iy\lb^{-\ell-k}k\bigl(M_i^{k-1}L_i^{k-1+\ell}\bigr),
$$
where $\dt(\lb,z):=\sum_{-\iy}^\iy\lb^{-n}z^{n-1}$.

A generating function of the {\bf symmetries on the $\tau$-manifold} is given by a
vector of 
 the vertex operators, based on the one of Date, Jimbo, Kashiwara and Miwa, acting
on a single
$\tau$-function,

$$
\begin{array}{ll}
X(t,\lb,\mu)&:=\exp\Bigl(\sum_1^\iy t_i(\mu^i-\lb^i)\Bigr)
\exp\Bigl(\sum_1^\iy(\lb^{-i}-\mu^{-i}){1\over i}
{\pl\over\pl t_i}\Bigr),\\
\\[3mm]
&=\sum^{\iy}_{k=0}\frac{(\mu -\lb)^k}{k!}\sum^{\iy}_{\ell
=-\iy}\lb^{-\ell-k}W_{\ell}^{(k)},\quad\mbox{with
}W_{\ell}^{(0)}=\dt_{\ell 0}.
\end{array}\leqno(1.33)
$$
So, the $\tau$-manifold symmetries for the 2-Toda lattice can be
expressed as a vector of vertex operators $X(t,\lb,\mu)$ acting on
the vector of
$\tau$-functions\footnote{using
$\left(\frac{\mu}{\lb}\right)^\al=\sum_{k\ge 0}\left({\al\atop
k}\right)\left({\mu-\lb\over\lb}\right)^k$ and
$\MAT{1}\al\\k\mat =\frac{(\al)_k}{k!}$}:
$$
\BX(t,\lb,\mu):=\Bigl(\Bigl({\mu\over\lb}\Bigr)^nX(t,\lb,
\mu)\Bigr)_{n\in\BZ} =\Biggl(\sum_{k=0}^\iy{(\mu-\lb)^k\over k!}
\sum_{\ell=-\iy}^\iy \lb^{-\ell-k}W_{n,\ell}^{(k)}\Biggr)_{n\in\BZ}
\leqno(1.34)
$$
$$
\tilde\BX(s,\lb,\mu):=\Bigl(\Bigl({\lb\over\mu}\Bigr)^nX(s,\lb,
\mu)\Bigr)_{n\in\BZ}=
\Biggl(\sum_{k=0}^\iy{(\mu-\lb)^k\over k!}
\sum_{\ell=-\iy}^\iy \lb^{-\ell-k}\tilde W_{n,\ell}^{(k)}\Biggr)_{n\in\BZ}
\leqno(1.35)
$$
with
$$
W_{n,\ell}^{(k)}=\sum_{j=0}^k\MAT{1}n\\j\mat(k)_jW_{\ell}^{(k-j)}
\quad\mbox{and}\quad
\tilde W_{n,\ell}^{(k)}=W_{-n,\ell}^{(k)}\Bigl|_{t\rg s}.
\leqno(1.36)
$$
One easily computes (remember the definition of the customary Virasoro
generators $J$'s in footnote 2)
$$
W_n^{(0)}=\dt_{n,0},\quad W_n^{(1)}=J_n^{(1)}\quad\mbox{and}\quad
W_n^{(2)}=J_n^{(2)}-(n+1)J_n^{(1)},\quad\quad n\in\BZ\leqno{(1.37)}
$$
and
$$
\begin{array}{lllllll}
W_{m,i}^{(1)}&=W_i^{(1)}+mW_i^{(0)}&&&&W_{m,i}^{(2)}&=W_i^{(2)}+2mW_i^{(1)}
+m(m-1)W_i^{(0)}\\
 &=J_i^{(1)}+m\dt_{i0}&&&& &=J_i^{(2)}+(2m-i-1)J_i^{(1)}
+m(m-1)\dt_{i0}.
\end{array}
\leqno(1.38)
$$

For future use, record the formulas

\medbreak
\noindent (1.39)
\begin{eqnarray*}
\frac{1}{2}W_{m,i}^{(2)}+(i+1)W_{m,i}^{(1)}&=&\frac{1}{2}J_i^{(2)}+
\Bigl(m+\frac{i+1}{2}\Bigr)
J_i^{(1)}+\frac{m(m+1)}{2}\dt_{i0}\\
\frac{1}{2}W_{m,i}^{(2)}-W_{m,i}^{(1)}&=&\frac{1}{2}J_i^{(2)}+
\Bigl(m-\frac{i+3}{2}\Bigr)
J_i^{(1)}+\frac{m(m-3)}{2}\dt_{i0}.
\end{eqnarray*}

\bigbreak

\noindent The corresponding expression $\tilde W^{(k)}_{m,i}$ can be read off from
the above, using (1.36), wich $J_n^{(k)}$ replaced by 
$\tilde J_n^{(k)} =
J_{n}^{(k)}\mid_{t\rg s}$.

The symmetry $\BY_N$ acting
on $\Psi$ and $L$ as in (1.30) and (1.31) lifts to the vertex operator
$\BX$ acting on $\tau$, as in (1.34) and (1.35), according to the 
Adler-Shiota-van Moerbeke formula [ASV2]:

\proclaim Theorem 1.2. The vector fields $\BY_{N_1}$ and $\BY_{N_2}$ acting on
the $\Psi$-manifold and the vertex operators of type
$\BX(t,\lb,\mu)$ and $\frac{\mu}{\lb}\tilde \BX(s,\lb,\mu)$ acting on the
$\tau$-manifold are related as
follows:
$$
{\BY_{(N_1,0)}\Psi\over\Psi}
= \left((e^{-\eta}-1) \frac{\BX(t,\lb,\mu)\tau}{\tau},
(\Lb e^{-\tilde\eta}-1){\BX(t,\lb,\mu)\tau\over\tau}
\right),
$$
$$
{\BY_{(0,N_2)}\Psi\over\Psi}
= {\mu\over\lb}\biggl((e^{-\eta}-1){\tilde\BX(s,\lb,\mu)\tau\over\tau},
(\Lb e^{-\tilde\eta}-1){\tilde\BX(s,\lb,\mu)\tau\over\tau}
\biggr).
$$
\bigskip

\noindent Expanding $\BX$ and $\tilde\BX$ in terms of  $W$-generators, and $N_i$ in
terms of $M_i^{\al} L_i^{\be}$, we obtain:
 
\proclaim Corollary 1.2.1. For $n,k\in\BZ$, $n\geq 0$,
the symmetry vector fields $\BY_{M_i^nL_i^{n+k}}$,\linebreak $(i=1,2)$ acting on
$\Psi$ lead to the correspondences
$$
-{((M_1^nL_1^{n+k})_{\ell}\Psi_1)_m\over\Psi_{1,m}z^m}
=
{1\over n+1}
(e^{-\eta}-1){W_{m,k}^{(n+1)}(\tau_m)\over\tau_m},\leqno{(1.40)}
$$
$$
{((M_1^nL_1^{n+k})_{u}\Psi_2)_m\over\Psi_{2,m}z^m}
=
{1\over n+1}\biggl(
e^{-\tilde\eta}{W_{m+1,k}^{(n+1)}(\tau_{m+1})\over\tau_{m+1}}
-{W_{m,k}^{(n+1)}(\tau_m)\over\tau_m}
\biggr),\leqno{(1.41)}
$$
$$
{((M_2^nL_2^{n+k})_{\ell}\Psi_1)_m\over\Psi_{1,m}z^m}
=
{1\over n+1}
(e^{-\eta}-1){\tilde W_{m-1,k}^{(n+1)}(\tau_m)
\over\tau_m},\leqno{(1.42)}
$$
$$
-{((M_2^nL_2^{n+k})_{u}\Psi_2)_m\over\Psi_{2,m}z^m}
=
{1\over n+1}\biggl(
e^{-\tilde\eta}{\tilde W_{m,k}^{(n+1)}(\tau_{m+1})\over\tau_{m+1}}
-{\tilde W_{m-1,k}^{(n+1)}(\tau_m)\over\tau_m}
\biggr),\leqno{(1.43)}
$$
and acting on $\Psi^*$
$$
\frac{(((M_1^nL_1^{n+k})_{\ell})^{\top}\Psi^*_1)_m}{\Psi^*_{1,m}z^{-m}}=\frac{1}{n+1}
(e^{\eta}-1)\frac{W^{(n+1)}_{m+1,k}(\tau_{m+1})}{\tau_{m+1}}
\leqno(1.44)
$$
$$
-\frac{(((M_1^nL_1^{n+k})_{u})^{\top}\Psi^*_2)_m}{\Psi^*_{2,m}z^{-m}}=\frac{1}{n+1}
\Biggl(e^{\tilde\eta}\frac{W^{(n+1)}_{m,k}(\tau_m)}{\tau_m}-
\frac{W^{(n+1)}_{m+1,k}(\tau_{m+1})}{\tau_{m+1}}\Biggr)
\leqno(1.45)
$$
$$
-\frac{(((M_2^nL_2^{n+k})_{\ell})^{\top}\Psi^*_1)_m}{\Psi^*_{1,m}z^{-m}}=\frac{1}{n+1}
(e^{\eta}-1)\frac{\tilde W^{(n+1)}_{m,k}(\tau_{m+1})}{\tau_{m+1}}
\leqno(1.46)
$$
$$
\frac{(((M_2^nL_2^{n+k})_{u})^{\top}
\Psi^*_2)_m}{\Psi^*_{2,m}z^{-m}}=\frac{1}{n+1}
\Biggl(e^{\tilde\eta}\frac{\tilde
W^{(n+1)}_{m-1,k}(\tau_m)}{\tau_m}-
\frac{\tilde W^{(n+1)}_{m,k}(\tau_{m+1})}{\tau_{m+1}}\Biggr).
\leqno(1.47) $$

\medbreak\noindent\underline{Remark}: The expansion of the vertex operator
(1.33) is obtained  by multiplying the series
$$
e^{\sum_1^{\iy}t_i(\mu^i-\lb^i)}=\sum_{k=0}^{\iy}(\mu-\lb)^k
\sum^{\iy}_{i=1}f_i^{(k)}(t)\lb^{i-1},
$$
and
$$
e^{\sum_1^{\iy}(\lb^{-i}-\mu^{-i})\frac{1}{i}\frac{\pl}{\pl t_i}}=
\sum^{\iy}_{k=0}(\mu-\lb)^k\sum^{\iy}_{i=0} g_i^{(k)}
(\frac{\pl}{\pl t})\lb^{-k-i},
$$
where $f_i^{(k)}(u)$ and $g_i^{(k)}(u)$ are polynomials in $u$, vanishing at $u=0$,
except $f_1^{(0)}=g_0^{(0)}=1$. Therefore
$$
X(t,\lb,\mu)=\sum^{\iy}_{k=0}\frac{(\mu-\lb)^k}{k!}\Bigl(
\sum^{-\iy}_{\ell
=-k}\lb^{-k-\ell}k!f_{-k-\ell +1}^{(k)}+\sum^{\iy}_{\ell =-\iy}
\lb^{-k-\ell} (\mbox{differential operator})\Bigr),
$$
showing, for all $\ell\in\BZ $,
$$
W_{\ell}^{(k)}=(\mbox{polynomial in }t, \mbox{without constant
term)} +\left\{\begin{array}{l}\mbox{pure
differential}\\\mbox{~~~~~operator}\end{array}\right\}
+\dt_{\ell,0}\dt_{k,0},\leqno{(1.48)}$$
with $$ W_{\ell}^{(k)}=(\mbox{pure differential operator) if and only if
}\ell\geq -k+1 \mbox{ with }\ell,k\neq 0. $$
Observe, in view of (1.36), that for $n,\ell\in\BZ$,
$$
\begin{array}{ll}
W_{n,\ell}^{(k)}&=(\mbox{differential operator)}+
\mbox{(polynomial in }t, \mbox{without}\\
& \quad\quad\quad\mbox{constant term})+(n)_k\dt_{\ell,0}\\
[3mm]
\tilde W_{n,\ell}^{(k)}&=(\mbox{differential operator)}+
\mbox{(polynomial in }s, \mbox{without}\\
& \quad\quad\quad\mbox{constant term})+(-n)_k\dt_{\ell,0}.
\end{array}\leqno(1.49)
$$

\section{A larger class of symmetries for special initial
conditions}

In this section we consider the submanifold of $L$, $M$ and $\Psi$ such
that polynomials $p(L_1,M_1,L_2,M_2)$ in $L_1,M_1,L_2,M_2$ are legitimate
objects; e.g., 
$L_1L_2$ makes sense. This point of view leads to a
much wider algebra of symmetries containing the ones of section 1. 
Henceforth, we change the point of view; namely, $p$ refers to a polynomial and not
to a vector of polynomials as in previous section.

From (1.11), it follows that, for such polynomials $p$, the Toda vector fields can
be rewritten as:
$$
{\pl p\over\pl t_n}=\left[\left(L_1^n\right)_u,p\right]\quad\hbox{and}\quad
{\pl p\over\pl s_n}=\left[\left(L_2^n\right)_{\ell},p\right]\quad n=1,2,\dots.
\leqno(2.1)
$$
Letting $p=p(L_1,M_1,L_2,M_2)$ be a
polynomial, we redefine symmetries as follows
$$
\BY_p\Psi=(-p_{\ell}\Psi_1,p_u\Psi_2)=
\left\{
\begin{tabular}{l}
$-(p,0)_-\Psi
$\\
$(0,p)_-\Psi
$
\end{tabular}
\right.\leqno{(2.2)}
$$
and thus
$$
\BY_p(L_1,L_2)=\left( \left[-p_{\ell},L_1 \right] , \left[ p_u,L_2   \right] 
\right)=\left\{
\begin{tabular}{l}
$[-(p,0)_-,L]
$\\
$[(0,p)_-,L]
$
\end{tabular}
\right.\leqno{(2.3)}
$$ and similarly for $L_1$, $L_2$ replaced by $M_1$, $M_2$.
Then
$$
\begin{array}{ll}
[\BY_p,\BY_{\bar p}]\Psi&=\Bigl((\BY_p(\bar p,0))_--(\BY_{\bar
p}(p,0))_-+[(p,0)_-,(\bar p,0)_-]\Bigr)\Psi\\
&=-\Bigl((\BY_p(0,\bar p))_--(\BY_{\bar
p}(0,p))_--[(0,p)_-,(0,\bar p)_-]\Bigr)\Psi\\
&=\Biggl(\Bigl(((\BY_p(\bar p))_{\ell}-(\BY_{\bar
p}(p))_{\ell}+[p_{\ell},\bar p_{\ell}]\Bigr)\Psi_1,\\
&\quad\quad - ((\BY_p(\bar p))_u-(\BY_{\bar p}(p))_u-[p_u,\bar
p_u]\Bigr)\Psi_2\Biggr)=\BY_{\{p,\bar p\}}\Psi
\end{array}\leqno(2.4)
$$
where
$$
\{p,\bar p\}:=-\BY_p(\bar p)+\BY_{\bar
p}(p)-[p_{\ell},\bar p_{\ell}]+[p_u,\bar p_u].
\leqno{(2.5)} $$

We verify that these more general vector fields are indeed
symmetries, i.e.
$$
(i)\qquad[\BY_p,\frac{\pl}{\pl t_n}]=0,\qquad \qquad(ii) \qquad
[\BY_p,\frac{\pl}{\pl s_n}]=0,$$
using the argument of (1.23).  
For the case (i), set
$$
Q=( q,0)=(L_1^n,0)   \quad P=(p,0),
 $$
and, using (2.1) and (2.2), we define
$$
\tilde\BZ_Q(P)=\frac{\pl}{\pl t_n}(p,0)=[Q_+,P] \mbox{  and 
} \tilde\BZ_Q\Psi=Q_+\Psi\leqno{{\rm (a)}} $$
$$
\tilde\BY_P(Q)=(-[p_{\ell}, q],0)=[-(p_{\ell},-p_u),(
q,0)]=[-P_-,Q], \mbox{  and  } \tilde\BY_P\Psi=-P_-\Psi; \leqno{{\rm (b)}} $$
we are exactly in the situation (1.21) and (1.23), leading at once to
$[\BY_P,\BZ_Q]=0$, which is (i).  A similar argument leads to (ii)

\medbreak

Let $\AR$ be the Lie algebra of generalized symmetries
of the  2-Toda vector fields, with the standard commutator $[ ~,~]$.
Also, the ring of polynomials $\BC[L_1,L_2$, $M_1,M_2]$ in the
(noncommutative) variables $L_1,L_2$, $M_1,M_2$, forms a Lie algebra, not
only with regard to the standard bracket $[~,~]$, but also with regard to the
new bracket $\{p,\bar p\}$ defined in (2.5),
as wil be seen in the next theorem; denote by
$$
A_1,B_1,C_1,...=\mbox{any monomial }(L_1,M_1,I)\mbox{   and   }
A_2,B_2,C_2,...=\mbox{any monomial }(L_2,M_2,I).
$$
 
\bigbreak

\proclaim Lemma 2.1. The two brackets $[~,~]$ and $\{~,~\}$ of the Lie algebra
$\BC[L_1,L_2$, $M_1,M_2]$ interact as follows: for monomials
 $p=A_1A_2B_1B_2C_1C_2\pp$ and $\bar p=\bar A_1 \bar A_2\bar B_1\bar B_2\bar
C_1\bar C_2\pp$ in the algebra, two different expressions hold,
$$
\begin{array}{ll}
\{p,\bar p\}&=[p,\bar p]-\bar A_1[p_1,\bar A_2]\bar B_1\bar
B_2\bar C_1\bar C_2\pp -\bar A_1\bar A_2\bar B_1[p,\bar B_2]\bar
C_1\bar C_2\pp -\pp\\
& \quad +A_1[\bar p,A_2]B_1B_2C_1C_2\pp +A_1A_2B_1[\bar
p,B_2]C_1C_2\pp +\pp\\
&=[p,\bar A_1]\bar A_2\bar B_1\bar B_2\bar C_1\bar C_2\pp +\bar
A_1\bar A_2[p,\bar B_1]\bar B_2\bar C_1\bar C_2\pp +\pp\\
& \quad +A_1[\bar p,A_2]B_1B_2C_1C_2\pp +A_1A_2B_1[\bar
p,B_2]C_1C_2\pp +\pp
\end{array}\leqno(2.6)
$$

\bigbreak
\proclaim Theorem 2.2.  There is a
 Lie algebra homomorphism 
$$
\BC[L_1,L_2,M_1,M_2],~ \{~\} \longmapsto \AR,~ [~] 
$$
$$
p\rg\BY_p:\BY_p\Psi=-(p,0)_-\Psi=(0,p)_-\Psi
$$
i.e.,
$$
[\BY_p,\BY_{\bar p}]\Psi=\BY_{\{p,\bar p\}}\Psi.
$$
Moreover, we have the following inclusion (Lie algebra (anti)-homomorphism)
$$
\AR\supset w_{\iy} \oplus w_{\iy} \mbox{   with   } \{A_1,A_2\}=0,
$$
and
$$
\{A_1,B_1\}=[A_1,B_1] \quad \{A_2,B_2\}=-[A_2,B_2];
$$
Also
$$
\AR\supset w_{\iy}\otimes\BC(h,h^{-1}),\quad\quad\mbox{(Loop algebra over
}w_{\iy}), \leqno{(2.7)}$$
in two different ways:
$$
\{A_1L_2^k,B_1L_2^{\ell}\}=[A_1,B_1]L_2^{k+\ell},\{L_1^kA_2,L_1^{\ell}B_2\}=
-L_1^{k+\ell}[A_2,B_2].\leqno{(2.8)}
$$

\proclaim Corollary 2.2.1. The vector fields
$$\BY_p\Psi=-(p,0)_-\Psi
\mbox{ with }p=M_1^nL_1^{n+k},~M_2^nL_2^{n+k},~L_1^{\al}L_2^{\be}$$
all commute
with the Toda vector fields $\pl / \pl t_n$ and $\pl / \pl s_n$. The
vector fields $\pl / \pl c_{\al,\be}$, corresponding to $p=L_1^{\al}L_2^{\be}$,
all commute among themselves.

\medbreak
\noindent\underline{Remark 2.2.2}: The $\tau$-function can then be considered as a
function of $t,s$ and $c$, as follows:
$$
\tau(c_{10}+t_1,c_{20}+t_2,...,;
c_{01}+s_1,c_{02}+s_2,...;c_{11},c_{12},c_{21},...).
$$

\medbreak

\noindent\underline{Proof of Lemma 2.1, Theorem 2.2 and Corollary 2.2.1.}:  Since
$\BY_p(.)$ and $[x,.]$ are derivations, one computes for $p=A_1A_2B_1B_2$ and $\bar
p=\bar A_1\bar A_2\bar B_1\bar B_2$, that
\begin{eqnarray*}
\BY_p(\bar p)&=&\BY_p(\bar A_1)\bar A_2\bar B_1\bar B_2+\bar
A_1\BY_p(\bar A_2)\bar B_1\bar B_2+\bar A_1\bar A_2\BY_p(\bar B_1)\bar
B_2+\bar A_1\bar A_2B_1\BY_p(\bar B_2)\\
&=&-[p_{\ell},\bar A_1]\bar A_2\bar B_1\bar B_2+\bar A_1[p_u,\bar
A_2]\bar B_1\bar B_2-\bar A_1\bar A_2[p_{\ell},\bar B_1]\bar
B_2+\bar A_1\bar A_2\bar B_1[p_u,\bar B_2]\\
&\mbox{\rm or }&
\left\{
\begin{array}{l}
=-[p_{\ell},\bar A_1\bar A_2\bar B_1\bar B_2]+\bar
A_1[p_{\ell}+p_u,\bar A_2]\bar B_1 \bar B_2+\bar A_1\bar A_2\bar
B_1[p_{\ell}+p_u,\bar B_2]\\ =[p_u,\bar A_1\bar A_2\bar B_1\bar
B_2]-[p_{\ell}+p_u,\bar A_1]\bar A_2\bar B_1\bar B_2
-\bar A_1\bar A_2[p_{\ell}+p_u,\bar B_1]\bar
B_2
\end{array} \right. \\
&\mbox{\rm or }&
\left\{
\begin{array}{l}
=-[\bar p_{\ell},\bar p]+\bar A_1[p,\bar A_2]\bar B_1\bar B_2+\bar
A_1\bar A_2\bar B_1[p,\bar B_2]\\
=[p_u,\bar p]-[p,\bar A_1]\bar A_2\bar B_1\bar B_2-\bar A_1\bar
A_2[p,\bar B_1]\bar B_2
\end{array}
\right.
\end{eqnarray*}
and by symmetry, using the first expression, one finds
$$
\BY_{\bar p}(p)=-[\bar p_{\ell},p]+A_1[\bar
p,A_2]B_1B_2+A_1A_2B_1[\bar p,B_2].
$$
Using the definition (2.5) of $\{p,\bar p\}$, one computes
\begin{eqnarray*}
\{p,\bar p\}&=&-\BY_p(\bar p)+\BY_{\bar p}(p)-[p_{\ell},\bar
p_{\ell}]+[p_u,\bar p_u]\\
&\mbox{\rm or }& \left\{
\begin{array}{l}
=[p_{\ell},\bar p]+[p,\bar p_{\ell}]-[p_{\ell},\bar p_{\ell}]+[p_u,\bar
p_u] \\
\quad -\bar A_1[p,\bar A_2]\bar B_1\bar B_2-\bar A_1\bar A_2\bar
B_1[p,\bar B_2]+A_1[\bar p,A_2]B_1B_2+A_1A_2B_1[\bar p,B_2]\\
=-[p_u,\bar p]+[p,\bar p_{\ell}]-[p_{\ell},\bar
p_{\ell}]+[p_u,\bar p_u]\\
\quad +[p,\bar A_1]\bar A_2\bar B_1\bar B_2+\bar A_1\bar
A_2[p,\bar B_1]\bar B_2+A_1[\bar p,A_2]B_1B_2+A_1A_2B_1[\bar
p,B_2]\end{array} \right. \\ &\mbox{\rm or }& \left\{
\begin{array}{l}
=[p,\bar p]-\bar A_1[p,\bar A_2]\bar B_1\bar B_2+A_1[\bar
p,A_2]B_1B_2-\bar A_1\bar A_2\bar B_1[p,\bar B_2]+A_1A_2B_1[\bar
p,B_2]\\
=[p,\bar A_1]\bar A_2\bar B_1\bar B_2+\bar A_1\bar A_2[p,\bar
B_1]\bar B_2+A_1[\bar p,A_2]B_1B_2+A_1A_2B_1[\bar p,B_2],
\end{array}
\right. \end{eqnarray*}
using
$$
[p_{\ell},\bar p]+[p,\bar p_{\ell}]-[p_{\ell},\bar p_{\ell}]+[p_u,\bar
p_u]=[p_{\ell},\bar p]+[p_u,\bar p_{\ell}]+[p_u,\bar p_u]=[p_{\ell},\bar
p]+[p_u,\bar p]=[p,\bar p]$$
and
$$
[p_u,\bar p]-[p,\bar p_{\ell}]+[p_{\ell},\bar p_{\ell}]-[p_u,\bar p_u]=0,
$$
establishing (2.6).

In particular we have the following:
$$
\{A_1A_2,B_1B_2\}=[A_1,B_1]A_2B_2-A_1B_1[A_2,B_2]\leqno{(2.9)}
$$
$$
\quad\quad\quad\{A_2A_1,B_2B_1\}=A_2B_2[A_1,B_1]-[A_2,B_2]A_1B_1
$$
$$
\{A_1A_2,B_2B_1\}=[A_1,B_2][B_1,A_2]-A_1[A_2,B_2]B_1+B_2[A_1,B_1]A_2,
$$
from which it follows at once that
$$
\{A_1,A_2\}=0,\quad\quad\{A_1,B_1\}=[A_1,B_1],\quad\quad\{A_2,B_2\}=-[A_2,
B_2];\leqno(2.10)
$$
this implies that $w_{\iy}\oplus w_{\iy}\subset\AR.$ Moreover, the first relation
of (2.9) leads to (2.8), implying (2.7) with $h=L_1$ or $h=L_2$, ending the
proof of theorem 2.2. From (2.9), it follows that
\medbreak
\noindent (2.11)
\begin{eqnarray*}
\{L_1^iL_2^j,L_1^{i'}L_2^{j'}\}&=&\{L_2^jL_1^i,L_2^{j'}
L_1^{i'}\}=0\\
\{L_1^iL_2^j,L_2^{i'}L_1^{j'}\}&=&[L_1^i,L_2^{i'}][L_1^{j'},L_2^j],
\end{eqnarray*}
establishing corollary 2.2.1.

\medbreak

\proclaim Theorem 2.3. Given functions
$$
f(L_1,L_2)=\sum a_{ij}L_1^iL_2^j\mbox{\,\,and\,\, }g(L_1,L_2)=\sum
b_{ij}L_1^iL_2^j, $$
then the following matrices form a Virasoro algebra (without central extension) for
the $\{\,,\,\}$-bracket\footnote{$f=f(L_1,L_2)=\sum_{i,j}a_{ij}L_1^iL_2^j$ and
$g=g(L_1,L_2)=\sum_{i,j}b_{ij}L_1^iL_2^j$}:
$$
\{L_1^{i+1}(M_1+f),L_1^{j+1}(M_1+f)\}=(i-j)L_1^{i+j+1}(M_1+f)
\leqno{(2.12)}
$$
$$
\{(M_2-g)L_2^{i+1},(M_2-g)L_2^{j+1})\}=(j-i)(M_2-g)L_2^{i+j+1}.
\leqno{(2.13)}
$$
If $\frac{\pl g}{\pl z_1}=\frac{\pl f}{\pl z_2}$, then the two
representations decouple $$
\{L_1^{i+1}(M_1+f),(M_2-g)L_2^{j+1}\}=0.\leqno{(2.14)}
$$

\medbreak

\noindent\underline{Proof}: At first, we need to prove
\medbreak
\noindent (2.15)
\begin{eqnarray*}
\{L_1^kM_1,L_1^{\al}fL_2^{\be}\}&=&-\al L_1^{k+\al
-1}fL_2^{\be}-L_1^{k+\al}\frac{\pl f}{\pl
z_1}L_2^{\be}\\ \{M_2L_2^k,L_1^{\al}fL_2^{\be}\}&=&\be
L_1^{\al }fL_2^{k+\be -1}+L_1^{\al}\frac{\pl f}{\pl z_2}L_2^{k+\be}.
\end{eqnarray*} Indeed, it suffices to do so for a monomial
$f=L_1^iL_2^j$; e.g., using (2.9), we have
\begin{eqnarray*}
\{L_1^kM_1,L_1^{\al +i}L_2^{\be +j}\}&=&[L_1^kM_1,L_1^{\al +i}]L_2^{\be
+j}=-(\al +i)L_1^{\al +i+k-1}L_2^{\be +j}\\
\{M_2L_2^k,L_1^{\al +i}L_2^{\be +j}\}&=&-L_1^{\al
+i}[M_2L_2^k,L_2^{\be +j}]=(\be +j)L_1^{\al +i}L_2^{\be
+j+k-1}.
\end{eqnarray*}
Using the relation (2.10), (2.11) and (2.15) above, we find
\medbreak
\noindent $\{L_1^{i+1}(M_1+f),L_1^{j+1}(M_1+f)\}$
\begin{eqnarray*}
&=&\{L_1^{i+1}M_1,L_1^{j+1}M_1\}+\{L_1^{i+1}f,L_1^{j+1}M_1\}+\{L_1^{i+1}M_1,
L_1^{j+1}f\}+\{L_1^{i+1}f,L_1^{j+1}f\}\\
&=&(i-j)L_1^{i+j+1}M_1+(i+1)L_1^{i+j+1}f+L_1^{i+j+2}\frac{\pl f}{\pl z_1}-
(j+1)L_1^{i+j+1}f-L_1^{i+j+2}\frac{\pl f}{\pl z_1}\\
&=&(i-j)L_1^{i+j+1}(M_1+f),
\end{eqnarray*}
and the same relations yield
\medbreak
\noindent $\{(M_2-g)L_2^{i+1},(M_2-g)L_2^{j+1}\}$
\begin{eqnarray*}
&=&\{M_2L_2^{i+1},M_2L_2^{j+1}\}-\{gL_2^{i+1},M_2L_2^{j+1}\}-\{M_2L_2^{i+1},
gL_2^{j+1}\}+\{gL_2^{i+1},gL_2^{j+1}\}\\
&=&(j-i)M_2L_2^{i+j+1}+(i+1)gL_2^{i+j+1}+\frac{\pl g}{\pl z_2}L_2^{i+j+2}-
(j+1)gL_2^{i+j+1}-\frac{\pl g}{\pl z_2}L_2^{i+j+2}\\
&=&(j-i)(M_2-g)L_2^{i+j+1},
\end{eqnarray*}
confirming (2.12) and (2.13). Finally, one checks (2.14),
\medbreak
\noindent $\{L_1^{i+1}(M_1+f),(M_2-g)L_2^{j+1}\}$
\begin{eqnarray*}
&=&\{L_1^{i+1}M_1,M_2L_2^{j+1}\}+\{L_1^{i+1}f,M_2L_2^{j+1}\}-\{L_1^{i+1}M_1,
gL_2^{j+1}\}-\{L_1^{i+1}f,gL_2^{j+1}\}\\
&=&0-L_1^{i+1}\frac{\pl f}{\pl z_2}L_2^{j+1}+L_1^{i+1}\frac{\pl g}{\pl
z_1}L_2^{j+1}+0=-L_1^{i+1}\Bigl(\frac{\pl f}{\pl z_2}-
\frac{\pl g}{\pl z_1}\Bigr)L_2^{j+1},
\end{eqnarray*}
ending the proof of Theorem 2.3.

In the next theorem, we provide the explicit solution to a large
class of symmetries and, in particular, to the system of differential
equations (0.3'). Since the $S_1$ and $S_2$ of the Borel decomposition of the
moment matrix $m_{\iy}$ will turn out to be the $S_1$ and $S_2$ of this
section, it will suffice to prove theorem 2.4. Note that symmetry vector fields
of the form
$$
\frac{\pl\Psi}{\pl
c}:=\BY_p\Psi=-(p,0)_-\Psi=(-p_{\ell},p_u)(\Psi_1,\Psi_2)
\mbox{\,\,with\,\,}p=p_1(L_1,M_1)p_2(L_2,M_2)\leqno{(2.16)}
$$
induce vector fields on $S_1^{-1}S_2$:
$$
\frac{\pl S_1^{-1}S_2}{\pl c}=p_1(\Lb,\varepsilon+\sum^{\iy}_{k=1}
kt_k\Lb^{k-1})S_1^{-1}S_2p_2(\Lb^{\top},\varepsilon^*+\sum^{\iy}_{k=1}
ks_k\Lb^{\top k-1}).\leqno{(2.17)}
$$

\proclaim Theorem 2.4. For symmetries of the form (2.16), we find
the solution

\medbreak\noindent (2.18) $\quad(S_1^{-1}S_2)(t,s,c)=$
$$
\sum_{r=0}^{\iy}\frac{c^r}{r!}p_1^r\Bigl(\Lb,\vr+
\sum_1^{\iy}kt_k\Lb^{k-1})e^{\sum_1^{\iy}t_k\Lb^k}(S_1^{-1}S_2)(0,0,0)
e^{-\sum_1^{\iy}s_k\Lb^{\top k}}p^r_2(\Lb^{\top},
\vr^*+\sum_1^{\iy}kt_k\Lb^{\top
k-1})
$$
with
$$
\tau_N(t,s,c)=\det\Bigl(S_1^{-1}S_2(t,s,c)_{ij}\Bigr)_{0\leq i,j\leq
N-1}.\leqno{(2.19)}
$$

\proclaim Corollary 2.4.1. In particular, symmetries of the form
$$
\frac{\pl\Psi}{\pl c_{\al\be}}:=\BY_{L_1^{\al}L_2^{\be}}\Psi,
$$
have the following solution
$$
S_1^{-1}S_2(t,s,c_{\al\be})=\sum^{\iy}_{r=0}
\frac{c^r_{\al\be}}{r!}\Lb^{r\al}
e^{\sum^{\iy}_1t_k\Lb^k}S_1^{-1}S_2(0,0,0)e^{-
\sum^{\iy}_1s_k\Lb^{\top k}}(\Lb^{\top})^{r\be}
$$

\proclaim Corollary 2.4.2. In general, setting $t=c_{10}$ and
$s=-c_{0,1}$, we have

\medbreak\noindent $S_1^{-1}S_2(t,s,c_{11},c_{12},\ldots)=$
$$
\renewcommand{\arraystretch}{0.5}
\begin{array}[t]{c}
\sum\\
{\scriptstyle (r_{\al\be})_{\al,\be\geq 0}\in\BZ^{\iy}}\\
{\scriptstyle \,\,\,\,\,_{(\al,\be)\neq (0,0)}}
\end{array}
\renewcommand{\arraystretch}{1}
\left(\prod_{(\al,\be)}\frac{c_{\al\be}^{r_{\al\be}}}{r_{\al\be}!}\right)
\Lambda^{\sum_{\al\geq 1}\al
r_{\al\be}}S_1^{-1}S_2(0,0,0,\ldots)\Lambda^{\top\sum_{\be\geq
1}\be r_{\al\be}};
$$

\medbreak

\noindent\underline{Proof of theorem 2.4}: Note that the vector
field (2.16) induces the vector field
$$
\BY_p(S_1)=-p_{\ell}S_1\mbox{\,\,and\,\,}\BY_p(S_2)=p_uS_2
$$
and thus for $p=p_1(L_1,M_1)p_2(L_2,M_2)$ as in (2.16), we find
\begin{eqnarray*}
\BY_p(S_1^{-1}S_2)&=&-S_1^{-1}(-p_{\ell}S_1)S_1^{-1}S_2
+S_1^{-1}p_uS_2=S_1^{-1}pS_2\\
&=&S_1^{-1}pS_2\\
&=&p_1(\Lb,\vr+\sum_1^{\iy}kt_k\Lb^{k-1})S_1^{-1}S_2p_2(\Lb^{\top},
\vr^*+\sum_1^{\iy} ks_k(\Lb^{\top})^{k-1}),
\end{eqnarray*}
a linear differential equation for $S_1^{-1}S_2$. Since for any linear
differential equation
$$
\frac{\pl x}{\pl c}=L(x)\quad\Longleftrightarrow
x(c)=e^{cL}x(0)=\sum\frac{c^r}{r!}L^rx(0),
$$
the conclusion (2.18) follows. Corollary 2.4.1. is merely a
special case of (2.18), while corollary 2.4.2 takes into account
the $t$-, $s$- and $c_{\al\be}$-flows all at once.

\section{Borel decomposition of moment matrices, $\tau$-functions and
string-orthogonal polynomials}

Consider the weight  $\rho (y,z) dy dz=e^{V(y,z)}dy dz$ on $\BR^2 $,
the corresponding inner
product $\la~,~\ra $ and the moment matrix\footnote{In this and
subsequent sections
$$
(t,s,c)=\Bigl((t_n)_{n\geq 1},(s_n)_{n\geq
1},(c_{\al\be})_{\al,\be\geq 1}\Bigr)=:
(c_{\al\be})_{\begin{array}{l}
\scriptsize \al,\be\geq 0\\
\scriptsize (\al,\be)\neq(0,0)
\end{array}}			
$$
with the understanding that $t_n=c_{n0}$, $s_n=-c_{0n}$.} $m_n(t,s,c)$, as in
(0.1),(0.2) and (0.3). In the introduction, 
{\it string-orthogonal polynomials} were defined as two sequences of monic
polynomials of degree
$i$, each depending on one variable ($y$ and
$z \in \BR$)
$$\{ p^{(1)}_i (y) \}^{\infty}_{i=0} \quad\mbox{and}\quad \{ p^{(2)}_i (z) \}
^{\infty}_{i=0},$$
orthogonal in the following sense
$$\la p^{(1)}_i ,p^{(2)}_j \ra = h_i \delta_{ij} .$$
Also consider the $N\times\iy$ matrix of Schur polynomials\footnote{The Schur
polynomials $p_i$, defined by
$e^{\sum^{\infty}_1 t_i z^i }=\sum^{\infty}_0 p_k(t)z^k$ and $p_k(t)=0$ for $k<0$,
are not to be confused with the string-orthogonal polynomials $p_i^{(k)}$, $k=1,2$.}
$p_i(t)$:
\begin{eqnarray*}
E_N(t)&=&\left(\begin{array}{ccccc|cc}
1&p_1(t)&p_2(t)&\pp&p_{N-1}(t)&p_N(t)&\pp\\
0&1&p_1(t)&\pp&p_{N-2}(t)&p_{N-1}(t)&\pp\\
\vdots&\vdots& & & \vdots&\vdots& \\
0&0&0&\pp&p_1(t)&p_2(t)&\pp\\
0&0&0&\pp&1&p_1(t)&\pp
\end{array}\right)	
\\
&=&(p_{j-i}(t))_{\begin{array}{l}
\scriptsize0\leq i\leq N-1\\
\scriptsize0\leq j<\iy
\end{array}}																																															
\end{eqnarray*}
The polynomials $p_n^{(1)}(y)$ and $p_n^{(2)}(z)$
have explicit expressions in terms of the moments
$$
p_n^{(1)}(y)=\frac{1}{\det m_n(t,s,c)}\det
\left(\begin{array}{ccc|c}
 & & &1\\
 & & &y \\
 &m_n(t,s,c)& &\vdots\\
 & & & \\
\cline{1-3}
\mu_{n,0}(t,s,c)&\pp&\mu_{n,n-1}(t,s,c)&y^n
\end{array}\right)
\leqno{(3.1)}
$$
and
$$
p_n^{(2)}(z)=\frac{1}{\det m_n(t,s,c)}\det
\left(\begin{array}{ccc|c}
 & & &1\\
 & & &z\\
 &m_n^{\top}(t,s,c)& &\vdots\\
 & & & \\
\cline{1-3}
\mu_{0,n}(t,s,c)&\pp&\mu_{n-1,n}(t,s,c)&z^n
\end{array}\right)\,,
$$
from which
$$
h_i=\la p_i^{(1)},p_i^{(2)}\ra =\la p_i^{(1)},z^i\ra =
\frac{\det m_{i+1}}{\det m_i}.
$$

For future use, we need some notation: for an arbitrary matrix
$$A = (a_{ij})_{0 \leq i,j \leq N - 1}$$
with $N$ finite or infinite, denote by $A_i $ the upper-left $i \times i$ minor
and $\Delta_i := \det A_i $,  with $i \geq 1$ and $\Delta_0=1$; assuming all
$\Delta_i \neq 0$, define the following monic polynomials of degree $i$ 
 ($0 \leq i
\leq N-1$)
$$
P_i (z,A) := \frac{1}{\Delta_i } \det
\left( 
\begin{array}{lllll}
 & & & &1\\
 &A_i & & &\vdots\\
 &    & & &z^{i-1} \\
a_{i0}&\ldots &a_{i,i-1} & &z^{i}
\end{array}
\right)
=: \sum_{0 \leq j \leq i} \frac{\Delta^{(i+1)}_{ji}}{\Delta_i } z^j 
= z^i +
\ldots , \leqno{(3.2)}
$$
with
$$
 \Delta^{(i+1)}_{ii} = \Delta_i  \mbox{ and }
\Delta^{(i+1)}_{ji} = 0, \quad i < j. 
$$
Given the character $\bar\chi(z) = (1,z,z^2 ,\ldots)^{\top} $, 
the lower triangular matrix
$S(A)$, with 1's along the diagonal, is defined such that
$$
 S(A) \chi (z) := (P_0 (z,A), P_1 (z,A), \ldots)^{\top}.\leqno{(3.3)}
$$
Also define the $N \times N$ diagonal matrix $h = h (A)$
$$
 h := diag (h_0 ,h_1 , \ldots,h_{N-1}):= diag
(\frac{\Delta_1
}{\Delta_0 } ,\ldots, \frac{\Delta_N }{\Delta_{N-1 }}
),\leqno{(3.4)}
$$
and so $\Dt_N=\prod_0^{N-1}h_i$. Theorem 3.1 shows that performing the Borel
decomposition of the matrix $m_{\iy}$ is tantamount to the construction
of the
string-orthogonal polynomials  . Given $m_{\iy}$, define $S(m_{\iy})$,
$S(m_{\iy}^{\top})$ and $h(m_{\iy})$ by means of (3.3) and (3.4). We now state
Theorem 3.1, whose proof can be found in [AvM3].

\proclaim Theorem 3.1.  The vectors of string-orthogonal polynomials are given
by $$p^{(1)} (y) = L^{-1} \chi (y) \mbox{ and } p^{(2)} (z) = (U^{-1})^{\top}
\chi (z)\leqno{(3.5)}$$
where $L$ and $U$ are the lower- and upper-triangular matrices respectively,
with $1$'s on the diagonal, appearing in the Borel decomposition of the
matrix $m_{\iy}$:
$$
m_{\iy} = L~ h~ U := S^{-1} (m_{\iy}) h (m_{\iy}) (S^{-1}( m_{\iy}^{\top} ))^{\top}.
\leqno{(3.6)}
$$

\proclaim Corollary 3.1.1. The string-orthogonality relations are equivalent to
the Borel decomposition of the moment matrix $m_{\iy}$: 
$$\la p^{(1)}_i , p^{(2)}_j \ra= h_i \delta_{ij} \Longleftrightarrow S(m_{\iy}) m_{\iy}
(S(m_{\iy}^{\top} ))^{\top} = h(m_{\iy}).\leqno{(3.7)}$$ 

\medbreak\noindent\underline{Remark 3.1.1}: Any matrix $\Dt$
(finite or infinite), with all $\Dt_i\neq 0$, can be realized as
a moment matrix $m_{\iy}$ and the recipe (3.6) provides its Borel
decomposition.

\vspace{1cm}
The determinant of the moment matrix  $m_N$ can be expressed in terms of a double
matrix integral, as stated in the following theorem:

\proclaim Theorem 3.2. The following
integral
\medbreak
\noindent {\rm (3.8)}
\begin{eqnarray*}
\tau_N(t,s,c)&:=&\int\int_{\vec{u},\vec{v}\in\BR^N}
e^{\sum^N_{k=1}V(u_k,v_k)}\prod_{i<j}(u_i-u_j)\prod(v_i-v_j)
\vec{du}\vec{dv}\\ &=&N!\det m_N(t,s,c)\\
&=&N!\det\Bigl(E_N(t)m_{\iy}(0,0,c)E_N(-s)^{\top}\Bigr)
 \end{eqnarray*}
is a $\tau$-function of $t$ and $-s$; in particular $\tau_N$ satisfies the
KP-hierarchy in $t$ and $s$, for each $N=0,1,...$\,; moreover the $h_n=\la
p_n^{(1)},p_n^{(2)}\ra$ are given by
$$
h_n(t,s,c)=\frac{\tau_{n+1}(t,s,c)}{\tau_n(t,s,c)}\quad\quad\quad
\tau_0(t,s,c)=1.\leqno{(3.9)} $$

\medbreak
\noindent\underline{Remark}: For $V(y,z)=\sum_1^{\iy}t_iy^i+cyz-\sum_1^{\iy}
s_iz^i$, the  integral (3.8) is well known [AS1] to be the two matrix 
integral, integrated over the space of Hermitean matrices of size $N$:
$$
\Bigl(\frac{\pi}{\sqrt{c}}\Bigr)^{-N^2+N}\cdot\prod^N_{k=1}k!
\int\int_{\HR_N\times\HR_N}dM_1dM_2e^{{\rm
tr}\,V(M_1,M_2)}.
$$
Before giving the proof of this theorem,
we shall need several Lemmas:

\proclaim Lemma 3.2.1.
$$
m_N(t,s)=E_N(t)m_{\iy}(0,0)E_N(-s)^{\top}.
$$
and
$$
m_{\iy}(t,s)=e^{\sum_1^{\iy} t_n \Lambda^n}m_{\iy}(0,0)e^{-\sum_1^{\iy} s_n
\Lambda^{\top n}} .
$$

\medbreak

\noindent\underline{Proof}:
\begin{eqnarray*}
m_N(t,s)&=&\Bigl(\int\int_{\BR^2}u^{\ell -1}v^{k-1}e^{\Sg
t_iu^i}e^{V_{12}(u,v)}e^{-\Sg s_jv^j}du\,dv\Bigr)_{1\leq\ell,k\leq
N}\\
&=&\Biggl(\sum^{\iy}_{i,j=0}p_i(t)\Bigl(\int\int_{\BR^2}
e^{V_{12}(u,v)}u^{i+\ell
-1}v^{j+k-1}du\,dv\Bigr)p_j(-s)\Biggr)_{1\leq\ell,k\leq N}\\
&=&\Bigl(\sum^{\iy}_{i,j=0}p_i(t)\mu_{i+\ell
-1,j+k-1}(0,0)p_j(-s)\Bigr)_{1\leq\ell,k\leq N}\\
&=&E_N(t)m_{\iy}(0,0)E_N(-s)^{\top}. \end{eqnarray*}
The second line follows from the first one, since $E_{\iy}(t)=e^{\sum t_n
\Lambda^n}$, establishing Lemma 3.2.1.

\proclaim Lemma 3.2.2. For arbitrary sequences of monic polynomials $p_k^{(1)}$ and
$p_k^{(2)}$, we have
\begin{eqnarray*}
& &\det(u_k^{\ell -1})_{1\leq\ell,k\leq N}
\det(v_k^{\ell -1})_{1\leq\ell,k\leq N} \\
& &\quad\quad\quad\quad=\sum_{\sg\in
S_N}\det\Bigl(u^{\ell-1}_{\sg(k)}
v^{k-1}_{\sg(k)}\Bigr)_{1\leq\ell,k\leq N}\\ &
&\quad\quad\quad\quad=\det\Bigl(p_{\ell-1}^{(1)}
(u_k)\Bigr)_{1\leq\ell,k\leq N}
\det\Bigl(p_{\ell-1}^{(2)}(v_k)\Bigr)_{1\leq\ell,k\leq N}\,.
\end{eqnarray*}

\medbreak

\noindent\underline{Proof}: This is a standard result about Vandermonde
determinants.

\vspace{1cm}

\noindent Consider a linear space
$W^0$ spanned by formal (Laurent) series in
$z$ (for large $z$):
$$
W^0=\mbox{{\rm
span}}\Bigl\{\sum^{N-1}_{j=-\iy}a_{jk}z^j,k=0,\pp,N-1\Bigr\}\oplus
z^NH_+.
$$

\proclaim Lemma 3.2.3. Letting
$$
W^t=e^{\Sg t_iz^i}W^0,
$$
the $\tau$-function
$$
\tau(t)=\det(\mbox{{\rm Proj}}:W^t\rightarrow H_+)
$$
has the form
$$
\tau(t)=\det(E_N(t)a_N),
$$
where $a_N$ is the $(\iy,N)$-matrix of coefficients
$$
a_N=\left(\begin{array}{ccc}
a_{N-1,0}&\pp&a_{N-1,N-1}\\
\vdots& &\vdots\\
a_{10}&\pp&a_{1,N-1}\\
a_{00}&\pp&a_{0,N-1}\\
\hline
a_{-10}&\pp&a_{-1,N-1}\\
a_{-20}&\pp&a_{-2,N-1}\\
\vdots& &\vdots
\end{array}\right)\leqno{(3.10)}
$$

\medbreak

\noindent\underline{Proof}:
\begin{eqnarray*}
W^t&=&\mbox{{\rm
span}}\Bigl\{\sum^{\iy}_{i=0}p_i(t)z^i\sum^{N-1}_{j=-\iy}a_{jn}
z^j,n=0,1,\pp,N-1\Bigr\}\oplus\Bigl\{\sum^{\iy}_0
p_i(t)z^{i+N+k},k=0,1,2,\pp\Bigr\}\\
&=&\mbox{{\rm
span}}\Bigl\{\sum_{k\in\BZ}z^k\sum_{j\geq\max(-k,-N+1)}p_{k+j}(t)
a_{-jn},n=0,1,\pp,N-1\Bigr\}\oplus\\
& &\hspace{7cm} \Bigl\{\sum^{\iy}_0
p_i(t)z^{i+N+k},k=0,1,2,\pp\Bigr\}.
\end{eqnarray*}
Therefore, recording the coefficients of $z^k$ in the $k$th row
starting from the bottom $(k=0,1,2,\pp)$, the projection $W^t\rg
H_+$ has the following matrix representation:
\begin{eqnarray*}
\tau(t)&=&\det\mbox{ {\rm
Proj}}(W^t\rg H_+)\\
&=&\det\left(\begin{array}{ccc|cccc}
 & & &\vdots&\vdots&\vdots& \\
 & & &p_2&p_1&1&\pp \\
\vdots& &\vdots&p_1&1&0&\pp \\
\renewcommand{\arraystretch}{0.5}
\begin{array}[t]{c}
\sum\\
{\scriptstyle j\geq -N+1}
\end{array}
\renewcommand{\arraystretch}{1}a_{-j0}p_{j+N}&
\pp&\renewcommand{\arraystretch}{0.5}
\begin{array}[t]{c}
\sum\\
{\scriptstyle j\geq -N+1}
\end{array}
\renewcommand{\arraystretch}{1}a_{-j,N-1}p_{j+N}&1&0&0&\pp \\
\hline
\renewcommand{\arraystretch}{0.5}
\begin{array}[t]{c}
\sum\\
{\scriptstyle j\geq -N+1}
\end{array}
\renewcommand{\arraystretch}{1}a_{-j0}p_{j+N-1}&\pp&\renewcommand{\arraystretch}{0.5}
\begin{array}[t]{c}
\sum\\
{\scriptstyle j\geq -N+1}
\end{array}
\renewcommand{\arraystretch}{1}a_{-j,N-1}p_{j+N-1}& & & & \\
\vdots& &\vdots& &O& & \\
\renewcommand{\arraystretch}{0.5}
\begin{array}[t]{c}
\sum\\
{\scriptstyle j\geq -1}
\end{array}
\renewcommand{\arraystretch}{1}a_{-j0}p_{j+1}&\pp&\renewcommand{\arraystretch}{0.5}
\begin{array}[t]{c}
\sum\\
{\scriptstyle j\geq -1}
\end{array}
\renewcommand{\arraystretch}{1}a_{-j,N-1}p_{j+1}& & & & \\
\renewcommand{\arraystretch}{0.5}
\begin{array}[t]{c}
\sum\\
{\scriptstyle j\geq 0}
\end{array}
\renewcommand{\arraystretch}{1}a_{-j0}p_j&\pp&\renewcommand{\arraystretch}{0.5}
\begin{array}[t]{c}
\sum\\
{\scriptstyle j\geq 0}
\end{array}
\renewcommand{\arraystretch}{1}a_{-j,N-1}p_j& & & &
\end{array}\right)\\
&=&\det\left(\begin{array}{ccc}
\renewcommand{\arraystretch}{0.5}
\begin{array}[t]{c}
\sum\\
{\scriptstyle j\geq -N+1}
\end{array}
\renewcommand{\arraystretch}{1}a_{-j0}p_{j+N-1}&\pp&\sum a_{-j,N+1}p_{j+N-1}\\
\vdots& &\vdots\\
\renewcommand{\arraystretch}{0.5}
\begin{array}[t]{c}
\sum\\
{\scriptstyle j\geq 0}
\end{array}
\renewcommand{\arraystretch}{1}a_{-j0}p_j& &\renewcommand{\arraystretch}{0.5}
\begin{array}[t]{c}
\sum\\
{\scriptstyle j\geq 0}
\end{array}
\renewcommand{\arraystretch}{1}a_{-j,N-1}p_j
\end{array}\right)\\
&=&\det(E_Na_N),
\end{eqnarray*}
ending the proof of Lemma 3.3.

\medbreak

\noindent\underline{Proof of Theorem 3.2}: Indeed, using Lemma
3.2.2 and Lemma 3.2.1, we have
\medbreak
\noindent (3.11)
\begin{eqnarray*}
&
&\int\int_{u,v\in\BR^N}e^{\sum^N_{k=1}V(u_k,v_k)}\prod_{i<j}(u_i-u_j)\prod_{i<j}(v_i-v_j)du_1\pp
du_N\,dv_1\pp dv_N \\
&=&\int\int_{u,v\in\BR^N}\prod^N_{k=1}e^{V(u_k,v_k)}
\sum_{\sg\in
S_N}\det\Bigl(u_{\sg(k)}^{\ell-1}v_{\sg(k)}^{k-1}\Bigr)_{1\leq\ell,k\leq
N}du_1\pp du_N\,dv_1\pp dv_N\\
&=&\sum_{\sg\in S_N}\det\Bigl(\int\int_{\BR^2}e^{V(u,v)}u^{\ell -1}
v^{k-1}du\,dv\Bigr)_{1\leq\ell,k\leq N}\\
&=&N!\det(m_N(t,s))\\
&=&N!\det(E_N(t)m_{\iy}(0,0)E_N(-s)^{\top}).
\end{eqnarray*}
Lemma 3.2.3 shows the latter expression is a $\tau$-function in $t$ and in $-s$,
where alternately $a_N^{(1)}(-s)=m_{\iy}(0,0)E^{\top}_N(-s)$ yields an 
initial plane in the Grassmannian parametrized by $s$, or
$a_N^{(2)}(t)=m_{\iy}^{\top}(0,0)E^{\top}_N(t)$ yields an 
initial plane parametrized by $t$. Finally (3.9) follows at once
from (3.1'), ending the proof of Theorem 3.2.

\section{From string-orthogonal polynomials to the two-Toda lattice and the
string equation}

Given the $(t,s,c)$-dependent weight $e^{V(y,z)}$ as in (0.1), its moment matrix
$m_{\iy}$ admits, according to Theorem 3.1,  the Borel decomposition
$$
m_{\iy}=S_1^{-1}h(\bar S_2^{-1})^{\top},\leqno{(4.1)}
$$
upon setting (in the notation (3.3)),
$$
S_1:=S(m_{\iy}),\quad\quad\bar S_2:=S(m_{\iy}^{\top}).\leqno{(4.2)}
$$
Also, according to Theorem 3.1, the semi-infinite matrices $S_1=S_1(t,s,c)$ and
$\bar S_2=\bar S_2(t,s,c)\in\DR_{-\iy,0}$ lead to string-orthogonal (monic)
polynomials
$(p^{(1)},p^{(2)})$ with $\la p_n^{(1)},p_m^{(2)}\ra =\dt_{n,m}h_n$; we also
define new matrices $\bar S_1,S_2\in\DR_{0,\iy}$ and vectors $\Psi_1$ and
$\Psi_2^*$, as follows:
$$ \begin{tabular}{lll}
$p^{(1)}(z)=:S_1\bar\chi(z)$&\mbox{and}&$p^{(2)}(z)=:\bar S_2\bar\chi(z)$\\ 
 & & \\
$\bar
S_1:=h(S_1^{-1})^{\top}$& \mbox{and}&$S_2:=h(\bar S_2^{-1})^{\top}$\\
 & & \\
$\Psi_1:=e^{\Sg t_kz^k}p^{(1)}(z)$&\mbox{and}&
$\Psi_2^*:=e^{-\Sg s_kz^{-k}}h^{-1}p^{(2)}(z^{-1}) $   \\
$\,\,\,~~~=e^{\Sg t_kz^k}S_1\bar\chi(z)$& &$\,\,\,~~~=e^{-\Sg
s_kz^{-k}}(S_2^{-1})^{\top}\bar\chi(z^{-1}).$   
\end{tabular}\leqno{(4.3)}
$$
Also define
matrices
$L_1,\bar L_2\in\DR_{-\iy,1}$,    $L_2,\bar L_1\in\DR_{-1,\iy}$, and $Q_1,\bar
Q_2\in\DR_{-\iy,-1}$ so that
$$
\begin{array}{lll}
{\rm (i)}&zp_n^{(1)}(z)=\sum_{\ell\leq
n+1}(L_1)_{n\ell}p_{\ell}^{(1)}(z)&
zp_n^{(2)}(z)=\sum_{\ell\leq n+1}(\bar
L_2)_{n\ell}p_{\ell}^{(2)}(z)\\
 & & \\
{\rm (ii)}&\frac{\pl}{\pl
z}p_n^{(1)}(z)=\sum_{\ell\leq
n-1}(Q_1)_{n\ell}p_{\ell}^{(1)}(z)&\frac{\pl}{\pl
z}p_n^{(2)}(z)=\sum_{\ell\leq n-1}(\bar
Q_2)_{n\ell}p_{\ell}^{(2)}(z)\\
 & & \\
{\rm (iii)}&\bar L_1=hL_1^{\top}h^{-1},&\bar L_2=hL_2^{\top}h^{-1}.
\end{array}\leqno(4.4)
$$
Using the latter, we check,
$$
z\Psi_1=e^{\Sg t_kz^k}zp^{(1)}(z)=e^{\Sg t_kz^k}L_1p^{(1)}=L_1\Psi_1,
$$
and
$$
z^{-1}\Psi_2^*(z)=e^{-\Sg s_kz^{-k}}h^{-1}z^{-1}p^{(2)}(z^{-1})=
e^{-\Sg s_kz^{-k}}h^{-1}\bar L_2p^{(2)}(z^{-1}))=
L_2^{\top}\Psi_2^*,
$$
leading to the formula (4.5) below upon setting $L^*_2=L_2^{\top}$; at the same time,
we also define the matrices $M_1$ and $M_2^*$:
$$(L_1,L_2^{\ast})(\Psi_1,\Psi_2^*)=(z,z^{-1})
(\Psi_1,\Psi_2^*)\leqno{(4.5)}$$
$$(M_1,M_2^*)(\Psi_1,\Psi_2^*):=\Bigl(\frac{\pl}{\pl
z},\frac{\pl}{\pl z^{-1}}\Bigr)(\Psi_1,\Psi_2^*) .
$$

\proclaim Theorem 4.1 (``String equations"). The matrices $L$ and
$Q$ satisfy the constraints\footnote{If
$f(y,z)=\Sg c_{ij}y^iz^j$, then $f(L_1,L_2)\equiv\Sg
c_{ij}L_1^iL_2^j$.},
$$
{\rm (i)}\quad Q_1+\frac{\pl V}{\pl y}(L_1,L_2)=0,\quad\quad {\rm
(ii)}\quad\bar Q_2+\left(
\frac{\pl V}{\pl z}(\bar L_1^{\top},\bar
L_2^{\top})\right)^{\top}=0,\leqno(4.6)
$$
and the matrices $L$ and $M$,
$$
{\rm (i)}\quad M_1+\frac{\pl V_{12}}{\pl y}(L_1,L_2)=0,\quad\quad {\rm
(ii)}\quad M_2^*+\Bigl(\frac{\pl V_{12}}{\pl
z}(L_1,L_2)\Bigr)^{\top}=0.\leqno{(4.7)} $$

\proclaim Corollary 4.1.1. If 
$$
V(y,z)=\sum_1^{\ell_1}t_iy^i+cyz
-\sum_1^{\ell_2}s_iz^i,
$$
then $L_1$ is a $\ell_2+1$-band matrix, having thus $\ell_2-1$ subdiagonals
below the main diagonal; also, $L_2$ is a $\ell_1+1$-band matrix, 
with $\ell_1-1$ subdiagonals above the main diagonal.

\proclaim Theorem 4.2 (``Toda equations"). The Borel decomposition of the moment
matrix for the weight $e^V$
$$
m_{\iy}=S_1^{-1}h(\bar S_2^{-1})^{\top}=S_1^{-1}S_2\leqno(4.8)
$$
provides matrices $S_i$ such that $L=(L_1,L_2)=(S_1\Lb
S_1^{-1},S_2\Lb^{\top}S_2^{-1})$. With regard to the parameters 
$(t,s,c)$ appearing
in the weight $e^V$, the matrices $m_{\iy}$ and $L$ satisfy the two-Toda equations
and symmetry equations
\footnote{Note $\frac{\pl L}{\pl
c_{\al\be}}=-[(L_1^{\al}L_2^{\be},0)_-,L]=[(0,L_1^{\al}L_2^{\be})_-,L]$.}
$$
\frac{\pl m_{\iy}}{\pl t_n}=\Lambda^nm_{\iy},\quad 
\frac{\pl m_{\iy}}{\pl s_n}=-m_{\iy}\Lambda^{\top n},\quad 
\frac{\pl m_{\iy}}{\pl c_{\al\be}}=\Lambda^{\al}m_{\iy}\Lambda^{\top\be},
$$
$$
\frac{\pl L}{\pl t_n}=[(L^n_1,0)_+,L]\quad \frac{\pl L}{\pl
s_n}=[(0,L_2^n)_+,L],\quad \frac{\pl L}{\pl
c_{\al,\be}}=-[(L_1^{\al}L_2^{\be},0)_-,L].\leqno(4.9)
$$

\proclaim Corollary 4.2.1. The moment matrix $m_{\iy}(t,s,c)$ 
can be expressed in
terms of its initial value $m_{\iy}(0,0,0)$ as follows:
$$
m_{\iy}(t,s,c)=\renewcommand{\arraystretch}{0.5}
\begin{array}[t]{c}
\sum\\
{\scriptstyle (r_{\al\be})_{\al,\be\geq 0}\in\BZ^{\iy}}\\
{\scriptstyle \,\,\,\,\,_{(\al,\be)\neq (0,0)}}
\end{array}
\renewcommand{\arraystretch}{1}\Biggl(\prod_{(\al,\be)}
\frac{c_{\al\be}^{r_{\al\be}}}{r_{\al\be}!}\Biggr)
\Lambda^{\sum_{\al\geq 1}\al r_{\al\be}}m_{\iy}(0,0,0)
\Lambda^{\top\sum_{\be\geq 1}\be r_{\al\be}}.
$$

\proclaim Corollary 4.2.2. The string-orthogonal polynomials,
defined in (3.1) or (4.3), have the following expression in terms
of the $\tau$-function $\tau_n=\det m_n$ (see (0.3) and footnote 2):
$$
p_n^{(1)}(y)=y^n\frac{\tau_n(t-[y^{-1}],s,c)}{\tau_n(t,s,c)}~~~~~~~~~~~~~~~~~~~~~
~~~~~~p_n^{(2)}(z)=z^n\frac{\tau_n(t,s+[z^{-1}],c)}{\tau_n(t,s,c)}.
$$
$$~~~~~~~~~~~~~~~~~=\sum_{0\leq k \leq n}
\frac{p_{n-k}(-\tilde\pl_t)\det m_n(t,s,c)}{\det m_n(t,s,c)}y^k
 ~~~~~~~~~~~~~~~~~~~~~~~=\sum_{0\leq k \leq n}
\frac{p_{n-k}(\tilde\pl_s)\det m_n(t,s,c)}{\det m_n(t,s,c)}z^k.
$$

\proclaim Theorem 4.3 (``Trace formula"). ([ASV4]) The following holds for
$n,m
\geq -1$:
$$
\sum_{0\leq i\leq
N-1}(L_1^{n+1}\,\,L_2^{m+1})_{ii}=\frac{1}{\tau_N(t,s)}p_{n+N}
(\tilde\pl_t)p_{m+N}(-\tilde\pl_s)\tau_1\circ\tau_{N-1}.
$$ 

\medbreak\noindent\underline{Remark 1}: Alternatively, also according to
[ASV4], the above formula can be written in terms of derivatives of
$\tau_{N-1}$ with coefficients involving the moments
$\mu_{ij}=\la y^i,z^j\ra$, defined in (0.3):
$$
\sum_{0\leq i\leq
N-1}(L_1^{n+1}\,\,L_2^{m+1})_{ii}=\frac{1}{\tau_N(t,s)}
\renewcommand{\arraystretch}{0.5}
\begin{array}[t]{c}
\sum\\
{\scriptstyle i+i'= n+N}\\
{\scriptstyle j+j'=m+N}\\
{\scriptstyle i,i',j,j'\geq 0}
\end{array}
\renewcommand{\arraystretch}{1}\mu_{ij}p_{i'}
(-\tilde\pl_t)p_{j'}(\tilde\pl_s)\tau_{N-1}(t,s).
$$

\vspace{0.5cm}
\noindent The reader will find the proof of Theorem 4.3 in [ASV4]. Before
proving Theorems 4.1 and 4.2, we first need  a few Lemmas:

\proclaim Lemma 4.4. The matrices $L_i,\bar L_i,M_1,M_2^*,Q_1,\bar Q_2$ satisfy:
$$
\left\{
\begin{array}{ll}
L_1=S_1\Lb S_1^{-1},&\bar  L_2=\bar S_2\Lb \bar S_2^{-1}\\
\bar L_1=\bar S_1\Lb^{\top}\bar S_1^{-1},&L_2=S_2\Lb^{\top}S_2^{-1}
\end{array}
\right.\leqno(4.10)
$$
and
$$
M_1=Q_1+\frac{\pl V_1}{\pl y}(L_1),\quad M_2^*=h^{-1}\bar
Q_2h+\frac{\pl V_2}{\pl z}(L_2^{\top}).\leqno(4.11)
$$

\medbreak

\noindent\underline{Proof}: From (4.3) and (4.4),
$$
L_1S_1\bar\chi(z)=L_1p_1=zp_1=zS_1\bar\chi(z)
=S_1z\bar\chi(z)=S_1\Lb\bar\chi(z),\quad\mbox{ for all } z,
$$
together with $\bar L_1=hL_1^{\top}h^{-1}$, implies the formulas (4.10)
for $L_1$ and $\bar L_1$; a similar argument implies those  for  
$L_2$ and $\bar L_2$.

To prove (4.11), observe, from (4.3) and (4.4), that for $\Psi_1=\Psi_1(t,s,c;z)$
\begin{eqnarray*}
M_1\Psi_1=\frac{\pl\Psi_1}{\pl z}
&=&\frac{\pl}{\pl z}\left(e^{\Sg t_kz^k}p^{(1)}(z)\right)\\
&=&\sum_{k\geq 1}kt_kz^{k-1}\Psi_1+e^{\Sg t_iz^k}\frac{\pl}{\pl z}p^{(1)}(z)\\
&=&(\Sg kt_kL_1^{k-1}+Q_1)\Psi_1,\\
&=&\Bigl(\frac{\pl
V_1}{\pl y}(L_1)+Q_1\Bigr)\Psi_1,
\end{eqnarray*}
and similarly that for $\Psi^*_2=\Psi_2^*(t,s,c;z)$,
\begin{eqnarray*}
M_2^{\ast}\Psi_2^{\ast}=\frac{\pl\Psi_2^*}{\pl z^{-1}}&=&\frac{\pl}{\pl
z^{-1}}\left(e^{-\Sg s_kz^{-k}}h^{-1}p^{(2)}(z^{-1})\right)\\
&=&-\Sg ks_k z^{-k+1}\Psi_2^*+e^{-\Sg s_kz^{-k}}h^{-1}\frac{\pl}{\pl
z^{-1}}p^{(2)}(z^{-1})\\
&=&-\Sg ks_k(L^{\top}_2)^{k-1}\Psi_2^*+e^{-\Sg s_kz^{-k}}h^{-1}
\bar Q_2p^{(2)}(z^{-1})\\
&=&\Bigl(\frac{\pl V_2(L_2^{\top})}{\pl
z}+h^{-1}\bar Q_2h\Bigr)\Psi_2^*, \end{eqnarray*}
concluding the proof of Lemma 4.4.

\medbreak

\noindent\underline{Proof of Theorem 4.1}: The following calculation, must be done
in the following spirit: for arbitrarily large $m$, set all $t_i=0$, for $i>2m$ and
$t_{2m}\neq 0$. Using in the fifth equality the fact that $c_{i0}=t_i$ and
$c_{0i}=-s_i$ (see (4.1)), compute, for all integer $n$ and $m\geq 0$, 
\begin{eqnarray*}
0&=&\int_{\BR}\frac{\pl}{\pl
y}\Bigl\{e^{V_1(y)}p_n^{(1)}(y)\Bigl(\int_{\BR}p_m^{(2)}
(z)e^{V(y,z)-V_1(y)}dz\Bigr)\Bigr\}dy\\
&=&\int_{\BR}e^{V_1(y)}\Bigl(\frac{\pl V_1}{\pl y}(y)p_n^{(1)}
(y)+\frac{\pl p_n^{(1)}}{\pl
y}(y)\Bigr)\Bigl(\int_{\BR}p^{(2)}_m(z)e^{V-V_1}dz\Bigr)dy\\
&&\hspace{3cm}+\int_{\BR}e^{V_1(y)}p_n^{(1)}(y)\Bigl(\int_{\BR}p_m^{(2)}(z)
e^{V-V_1}\frac{\pl}{\pl y}(V-V_1)dz\Bigr)dy\\
&=&\int_{\BR}\sum_{\ell}(Q_1)_{n\ell}p_{\ell}^{(1)}(y)\left(
   \int_{\BR}p_m^{(2)}(z)e^Vdz\right)dy
  +\int_{\BR}p_n^{(1)}(y)\left(\int_{\BR}p_m^{(2)}(z)e^V
   \frac{\pl V}{\pl y}dz\right)dy\\
&=&\sum_{\ell}(Q_1)_{n\ell}\int\int_{\BR^2}p^{(1)}_{\ell}
(y)p_m^{(2)}(z)e^{V(y,z)}dydz\\
&&\hspace{3cm}+\int\int_{\BR^2}p^{(1)}_n(y)p_m^{(2)}(z)
\renewcommand{\arraystretch}{0.5} \begin{array}[t]{c}
\sum\\
{\scriptstyle i\geq 0}\\
{\scriptstyle j\geq 0}
\end{array}
\renewcommand{\arraystretch}{1}ic_{ij}y^{i-1}z^je^Vdydz\\
&=&(Q_1)_{nm}h_m+\sum_{i,j}ic_{ij}\int\int_{\BR^2}(y^{i-1}
p_n^{(1)}(y))(z^jp_m^{(2)}(z))dydz\\
&=&(Q_1)_{nm}h_m+\sum_{i,j}ic_{ij}\int\int_{\BR^2}\Bigl(\sum_{\al}
(L_1^{i-1})_{n\al}p_{\al}^{(1)}(y)\Bigr)
\Bigl(\sum_{\be}
(\bar L_2^j)_{m\be}p_{\be}^{(2)}(z)\Bigr)e^Vdydz\\
&=&(Q_1)_{nm}h_m+\sum_{i,j}ic_{ij}\sum_{\al,\be}
(L_1^{i-1})_{n\al}(\bar L_2^j)_{m\be}\int\int_{\BR^2}p_{\al}^{(1)}
(y)p_{\be}^{(2)}(z)e^Vdydz\\
&=&(Q_1)_{nm}h_m+\sum_{i,j,\al\geq 0}ic_{ij}
(L_1^{i-1})_{n\al}(\bar L_2^j)_{m\al}h_{\al}\\
&\stackrel{(\ast)}{=}&(Q_1)_{nm}h_m+\sum_{i,j,\al\geq 0}ic_{ij}
(L_1^{i-1})_{n\al}(L_2^{j})_{\al m}h_m\\
&=&\left((Q_1)_{nm}+\sum_{i,j\geq 0}ic_{ij}(L_1^{i-1}L_2^j)_{nm}\right)h_m\\
&=&\left(Q_1+\frac{\pl V}{\pl y}(L_1,L_2)\right)_{nm}h_m,
\end{eqnarray*}
upon using, in the equality $\stackrel{(\ast)}{=}$, the property $h(\bar
L_2^j)^{\top}=L_2^jh$ (see (4.4, iii)); since $h_m\neq 0$, relation
(4.6, i) holds.

The proof of (4.6, ii) is similar to the previous
proof, upon interchanging $1\lrg 2$ and $m\lrg n$; indeed, compute for all $m$ and
$n$,
\begin{eqnarray*}
0&=&\int_{\BR}\frac{\pl}{\pl z}e^{V_2(z)}p_m^{(2)}(z)\Bigl(\int_{\BR}
p_n^{(1)}(y)e^{V(y,z)-V_2(z)}dy\Bigr)dz\\
&=&(\bar Q_2)_{mn}h_n+\sum_{i,j\geq
0}jc_{ij}\int\int_{\BR^2}(z^{j-1}p_m^{(2)}(z))(y^ip_n^{(1)}(y)
e^Vdydz\\
&=&(\bar Q_2)_{mn}h_n+\sum_{i,j}jc_{ij}\sum_{\al,\be}(\bar
L_2^{j-1})_{m\al}(L_1^i)_{n\be}\int\int
p_{\al}^2(z)p_{\be}^1(y)e^Vdydz\\
&=&(\bar Q_2)_{mn}h_n+\sum_{i,j,\al\geq 0}jc_{ij}(\bar
L_2^{j-1})_{m\al}(L_1^i)_{n\al}h_{\al}\\
&\stackrel{(\ast)}{=}&\Bigl((\bar Q_2)_{mn}+\sum_{i,j,\al\geq
0}jc_{ij}(\bar L_2^{j-1})_{m\al}(\bar L_1^i)_{\al
n}\Bigr)h_n\\
&=&\left(\bar Q_2+
\left(\frac{\pl V}{\pl z}(\bar L_1^{\top},\bar
L_2^{\top})\right)^{\top}\right)_{m,n} h_n,
\end{eqnarray*}
upon using in $\stackrel{(\ast)}{=}$ that
$h(\bar L_1^i)^{\top}=L_1^i h$ (see (4.5, iii)).

Now observe from (4.6)(i) and (4.11) that
\begin{eqnarray*}
0=Q_1+\frac{\pl V}{\pl y}(L_1,L_2)&=&Q_1+\frac{\pl V_1}{\pl
y}(L_1)+\frac{\pl V_{12}}{\pl y}(L_1,L_2)\\
&=&M_1+\frac{\pl V_{12}}{\pl y}(L_1,L_2),
 \end{eqnarray*}
yielding (4.7)(i). To prove (4.7)(ii), compute from (4.6)(ii)
and (4.11) that
\begin{eqnarray*}
0&=&h^{-1}\bar Q_2h+h^{-1}\left(
\frac{\pl V}{\pl z}(\bar L_1^{\top},\bar
L_2^{\top})\right)^{\top}h\\
&=&h^{-1}\bar Q_2h+\left(\frac{\pl V}{\pl z}(h\bar L_1^{\top}h^{-1},
h \bar L_2^{\top}h^{-1})\right)^{\top}\\
&=&h^{-1}\bar Q_2h+\left(\frac{\pl V}{\pl z}(L_1,L_2)\right)^{\top},\\
&=&h^{-1}\bar Q_2h+\frac{\pl V_2}{\pl z}(L_2^{\top})+
\Bigl(\frac{\pl V_{12}}{\pl z}(L_1,L_2)\Bigr)^{\top}\\
&=&M^*_2+\Bigl(\frac{\pl V_{12}}{\pl z}(L_1,L_2)\Bigr)^{\top},
\end{eqnarray*}
concluding the proof of Theorem 4.1.

\medbreak

\noindent\underline{Proof of Corollary 4.1.1}: From (4.6(i)) and (4.6(ii)), it
follows that
$$
Q_1+\sum^{\ell_1}_1it_iL_1^{i-1}+cL_2=0~~~\mbox{  and  }~~~
h\bar Q_2^{\top}h^{-1} +cL_1-\sum_1^{\ell_2}is_iL_2^{i-1}=0.
$$
But $Q_1$ is zero on and above the diagonal; $\sum^{\ell_1}_1it_iL_1^{i-1}$
has $\ell_1-1$ subdiagonals above the main diagonal and zeroes beyond, and thus
also $L_2$. Similarly $\bar Q_2^{\top}$has zeroes on and below the diagonal; 
$\sum_1^{\ell_2}is_iL_2^{i-1}$ has $\ell_2-1$ subdiagonals below the main
diagonal and zeroes beyond, and thus also $ L_1$, establishing the corollary.

\medbreak

\proclaim Lemma 4.5.  The wave operators $S=(S_1,S_2)$ satisfy for $n=1,2,...$,
and
$\al,\be=0,1,...$,
$$
  \vcenter{\ialign{$\displaystyle#$\hfil&& \qquad$\displaystyle#$\crcr
  \frac{\pl S_1}{\pl t_n}=-(L_1^n)_{\ell}S_1,
  &\frac{\pl S_1}{\pl s_n}=(L_2^n)_{\ell}S_1,
  &\frac{\pl S_1}{\pl c_{\al\be}}=-(L_1^{\al}L_2^{\be})_{\ell}S_1 
  \cr\noalign{\vskip3pt}
  \frac{\pl S_2}{\pl t_n}=(L_1^n)_uS_2,
  &\frac{\pl S_2}{\pl s_n}=-(L_2^n)_uS_2,
  &\frac{\pl S_2}{\pl c_{\al\be}}=(L_1^{\al}L_2^{\be})_uS_2\cr}}
  \leqno(4.12)$$
and the wave vectors $\Psi_1$ and $\Psi_2^{\ast}$
$$
  \everycr{\noalign{\vskip3pt}}
  \left\{\vcenter{\ialign{\hfil#\qquad&
    \hfil$\displaystyle#$& $\displaystyle{}#$\hfil\crcr
  (i)&
  \frac{\pl (\Psi_1, \Psi_2^{\ast})}{\pl t_n} 
  &=\left((L_1^n)_{u,} - ((L_1^n)_u)^{\top}\right)  (\Psi_1,\Psi_2^{\ast}) \quad
  n=1,2,\ldots 
  \cr
  (ii)&
  \frac{\pl (\Psi_1, \Psi_2^{\ast})}{\pl s_n} 
  &=\left((L_2^n)_{\ell,}-((L_2^n)_{\ell})^{\top}\right) (\Psi_1,\Psi_2^{\ast}) 
  \cr
  (iii)&
 \frac{\pl (\Psi_1, \Psi_2^{\ast})}{\pl c_{\al \be}} 
  &=\left( -{(L_1^{\al} L_2^{\be})}_{\ell ,}
     -((L_1^{\al}L_2^{\be})_u)^{\top}\right) (\Psi_1,\Psi_2^{\ast})
  \quad \al, \be \geq  0.\cr}}\right.
  \leqno{(4.13)}$$

\medbreak

\noindent\underline{Proof}: Since $p_k^{(1)}$ and $p_k^{(2)}$ are monic polynomials,
$$
\frac{\pl p_k^{(i)}}{\pl
c_{\al\be}}=\sum_{m< k}A_{km}^{(i)}p_m^{(i)}\quad i=1,2
\leqno(4.14)
$$
with $A^{(1)}$ and $A^{(2)}\in\DR_{-\iy,-1}$. So, for arbitrary $\ell$ and $k\geq 0$,
\begin{eqnarray*}
&\,&\frac{\pl}{\pl c_{ij}}\la p_k^{(1)},p_{\ell}^{(2)}\ra \\
&=&\la\frac{\pl p_k^{(1)}}{\pl c_{ij}},p_{\ell}^{(2)} \ra +
\la p_k^{(1)},\frac{\pl p_{\ell}^{(2)}}{\pl c_{ij}}\ra +
\la y^ip_k^{(1)},z^jp_{\ell}^{(2)}\ra \\
&=&\sum_{m<k}A^{(1)}_{km}\la p_m^{(1)},p_{\ell}^{(2)}\ra
+\sum_{m<\ell}A_{\ell m}^{(2)}\la p_k^{(1)},p^{(2)}_m\ra
+\sum_{m,n}(L_1^i)_{km}(\bar L^j_2)_{\ell n}\la
p_m^{(1)},p_n^{(2)}\ra \\
&=&A_{k\ell}^{(1)}h_{\ell}+A_{\ell
k}^{(2)}h_k+\sum_r(L_1^i)_{kr}(\bar L_2^j)_{\ell r}h_r
\end{eqnarray*}
$$
\mbox{or }\left\{
\begin{array}{l}
=A_{k\ell}^{(1)}h_{\ell}+A_{\ell
k}^{(2)}h_k+\renewcommand{\arraystretch}{0.5}
\begin{array}[t]{c}
\sum\\
{\scriptstyle r}
\end{array}
\renewcommand{\arraystretch}{1}
(L_1^i)_{kr}h_{\ell}(L_2^j)_{r\ell}\\
=A_{k\ell}^{(1)}h_{\ell}+A_{\ell k}^{(2)}h_k+\renewcommand{\arraystretch}{0.5}
\begin{array}[t]{c}
\sum\\
{\scriptstyle r}
\end{array}
\renewcommand{\arraystretch}{1}
(\bar
L_2^j)_{\ell r}(\bar L_1^i)_{rk}h_k
\end{array}
\right. \mbox{ using (4.4) (iii)}
$$
leading to
$$
  \left\{\,\vcenter{\ialign{$#$\hfil\qquad&
    $\displaystyle#$\hfil\crcr
  \ell <k&
  0=\frac{\pl}{\pl c_{ij}}
  \Bigl\la p_k^{(1)},p_{\ell}^{(2)}\Bigr\ra
  =\Bigl(A_{k\ell}^{(1)}+(L_1^iL_2^j)_{k\ell}\Bigr)h_{\ell}\cr
  \ell =k&
  \frac{\pl}{\pl c_{ij}}h=(L_1^iL_2^j)_0 h\cr
  \ell >k&
  0=\frac{\pl}{\pl c_{ij}}\la p_k^{(1)},p_{\ell}^{(2)}\ra 
  =\Bigl(A_{\ell k}^{(2)}+(\bar L_2^j\bar L_1^i)_{\ell k}\Bigr)h_k.
  \cr}}\right.
  \leqno(4.15)$$

Setting these expressions for the matrices $A^{(i)}\in\DR_{-\iy,-1}$ in (4.14), leads
to
$$
\frac{\pl p^{(1)}}{\pl
c_{\al\be}}=-(L_1^{\al}L_2^{\be})_{\ell}p^{(1)}\quad\mbox{ and }
\quad\frac{\pl p^{(2)}}{\pl
c_{\al\be}}=-(\bar L_2^{\be}\bar
L_1^{\al})_{\ell}p^{(2)}.\leqno{(4.16)} $$
Since $p^{(1)}=S_1\bar\chi$ and $p^{(2)}=\bar S_2\bar\chi$, 
(4.16) leads to
$$
\frac{\pl S_1}{\pl c_{\al\be}}=
-(L_1^{\al}L_2^{\be})_{\ell}S_1\quad\mbox{ and }\quad
\frac{\pl \bar S_2}{\pl
c_{\al\be}}=-(\bar L_2^{\be}\bar L_1^{\al})_{\ell}\bar
S_2,\leqno{(4.17)} $$
for integer $\al,\be\geq 0$, and since $S_2 = h(\bar S_2^{-1})^{\top},$
\begin{eqnarray*}
\frac{\pl S_2}{\pl c_{\al \be}} 
&=& \frac{\pl h}{\pl c_{\al \be}}\left( \bar S_2^{-1} \right)^{\top}
-h\left( \bar S_2^{-1} \frac{\pl \bar S_2}{\pl c_{\al \be}}      \bar S_2^{-1}
\right)^{\top} \\
&=&\left( (L_1^{\al} L_2^{\be})_0+h((\bar
L_2^{\be}\bar L_1^{\al})_{\ell})^{\top}h^{-1}   \right)S_2,
\quad\mbox{using (4.15), 
(4.17),}\\ &=&\left( (L_1^{\al}
L_2^{\be})_0+\Biggl(((L_1^{\al}L_2^{\be})^{\top})_{\ell}
\Biggr)^{\top}\right)S_2,\quad\mbox{using (4.4(iii))}\\
&=&(L_1^{\al} L_2^{\be})_u S_2, 
\end{eqnarray*} from which it follows that
 $$
\frac{\pl (S_2^{-1})^{\top}}{\pl c_{\al \be}} = -{((L_1^{\al}L_2^{\be}
)}_{u})^{\top} (S_2^{-1})^{\top}. \leqno{(4.18)}
$$
In particular, setting $c_{n0}=t_n$ or $-c_{0n}=s_n$ leads to (4.12).

Since, according to the definition (4.3),
$$
\Psi_1=e^{\Sg t_kz^k}S_1\bar\chi(z)\quad\mbox{and}\quad\Psi^*_2=e^{-\Sg
s_kz^{-k}}(S_2^{-1})^{\top}\bar\chi(z^{-1})
$$
depends on $c_{\al\be}$, $\al,\be\geq 1$, through $S_1$ and $S_2$ only, the
expressions for $\pl S_1/\pl c_{\al\be}$ and $\pl S_2/\pl c_{\al\be}$ yield the
corresponding derivatives (4.13, iii). Equations (4.13, i) and (4.13, ii) follow
from (4.13, iii), taking into account the fact that $t_n$ and $s_n$ also appear in
the exponents of $(\Psi_1,\Psi^*_2)$. This ends the proof of Lemma 4.5.

\medbreak

\noindent\underline{Proof of Theorem 4.2}: The formulas
$$
L_1=S_1\Lb S_1^{-1},\quad\quad L_2=S_2\Lb^{\top}S_2^{-1},
$$
and the expression (4.8) for $m_{\iy}$ combined with the differential equations
(4.12) for $S_i$ lead to
\begin{eqnarray*}
\frac{\pl m_{\iy}}{\pl c_{\al\be}}&=&\frac{\pl S_1^{-1}S_2}{\pl c_{\al\be}}\\
&=&S_1^{-1}(L_1^{\al}L_2^{\be})_{\ell}S_1^{-1}S_1S_2+S_1^{-1}
(L_1^{\al}L_2^{\be})_uS_2\\
&=&S_1^{-1}L_1^{\al}L_2^{\be}S_2\\
&=&\Lb^{\al}S_1^{-1}S_2\Lb^{\top\be}=\Lb^{\al}m_{\iy}\Lb^{\top\be},
\end{eqnarray*}
and thus the conclusion of Theorem 4.2. Corollary 4.2.1 is an
immediate consequence of Theorem 4.2, while Corollary 4.2.2
follows from (1.16), (1.17) and (4.3).

\section{Virasoro constraints on two-matrix integrals}

For the general 2-Toda lattice, define the bi-infinite matrices for 
all integer $i\geq -1$,
$$
v_i^{(1)} = L_1^{i+1} \bigl(M_1 + \frac{\pl V_{12}}{\pl y}(L_1,
L_2)\bigr),
\quad
v_i^{(2)} =-\bigl( M_2^{\ast \top} + \frac{\pl V_{12}}{\pl
z}(L_1,L_2)\bigr) L_2^{i+1},\leqno{(5.1)}
$$
\hfill (bi-infinite 2-Toda lattice).\newline Note that, according to
the formula preceeding (1.10), we have
$$
M_2^{\ast\top}=L_2^{-1}-M_2.\leqno{(5.2)}
$$
In view of formulas (1.39), define differential operators ${\cal
K}^{(\al)}_{m,i}$ and $\LR^{(\al)}_{m,i}$ for $m\geq 0$, $i\geq -1$,
$\al =1,2$,
\medbreak
\noindent (5.3)
\begin{eqnarray*}
\KR^{(1)}_{m,i}&:=&\frac{1}{2}W_{m,i}^{(2)}+(i+1)W_{n,i}^{(1)}+
\sum_{r,s\geq 1}rc_{rs}\frac{\pl}{\pl c_{i+r,s}}\\
&=&\frac{1}{2} J_i^{(2)} +
(m+\frac{i+1}{2})J_i^{(1)} +\sum_{r,s \geq 1} rc_{rs} \frac{\pl}{\pl
c_{i+r,s}}+\frac{m(m+1)}{2}\delta_{i0}\\
&=:&\LR^{(1)}_{m,i}+\frac{m(m+1)}{2}\delta_{i0}
\end{eqnarray*}
and
\medbreak
\noindent (5.4)
\begin{eqnarray*}
\KR^{(2)}_{m,i}&:=&\frac{1}{2}\tilde
W_{m-1,i}^{(2)}-\tilde W_{m-1,i}^{(1)}+
\sum_{r,s\geq 1}sc_{rs}\frac{\pl}{\pl c_{r,s+i}}\\
&=&\frac{1}{2} \tilde J_i^{(2)} -
(m+\frac{i+1}{2})\tilde J_i^{(1)} +\sum_{r,s \geq 1} sc_{rs}
\frac{\pl}{\pl c_{r,s+i}}+\frac{(m-1)(m+2)}{2}\delta_{i0}\\
&=:&\LR^{(2)}_{m,i}+\frac{(m-1)(m+2)}{2}\delta_{i0}.
\end{eqnarray*}
Consider the usual weight $e^V$, the corresponding
string-orthogonal polynomials, the semi-infinite wave vectors
$\Psi_1$ and $\Psi_2^*$ derived from them in (4.3) and the
semi-infinite matrices $L_1,L_2,M_1,M_2^*$ defined in (4.4) and
(4.5). Note that in this context neither $L_2^{-1}$ nor $M_2$
appearing in (5.2) make sense although the combination
$M_2^{\ast\top}=L_2^{-1}-M_2$ makes perfectly good sense.

Recall the bracket $\{\,,\,\}$ introduced in (2.5) and made
explicit in (2.6); we now state

\proclaim Theorem 5.1. For the general 2-Toda lattice, the
bi-infinite matrices
$v_i^{(\al)}$ ($i\geq -1$,
$\al =1,2$) form a (decoupled) algebra $Vir^+\otimes Vir^+$ for the
bracket
$\{\,,\,\}$:
$$
\{v_i^{(\al)},v_j^{(\al)}\}=(-1)^{\al+1}(i-j)v_{i+j}^{(\al)}\quad\quad\{v_i^{(1)},
v_j^{(2)}\}=0.\leqno{(5.5)} $$ For string-orthogonal polynomials, we
have
$$v_i^{(\al)}=0\quad\mbox{and}\quad \tau_0=1;$$ they imply for $$
\tau_m=\frac{1}{m!}\int\int_{u,v\in\BR^m}e^{\sum_1^mV(u_k,v_k)}
\prod_{i<j}(u_i-u_j)\prod_{i<j}(v_i-v_j)du\,dv
$$
the
algebra
$Vir^+\otimes Vir^+$ of constraints $$
\Bigl(\LR_{m,i}^{(\al)}+\frac{m(m+1)}{2}\dt_{i,0}\Bigr)\tau_m=0,\quad
\mbox{for }m\geq 0,i\geq -1,\al =1,2.\leqno{(5.6)} $$

\proclaim Lemma 5.2. For the general 2-Toda lattice, we have the
following correspondence for the vector field $\BY_{v_i^{(\al)}}$
acting on the wave functions $(\Psi_1,\Psi^*_2)$:
$$
\begin{array}{ll}
-\frac{((v_i^{(\al)})_{\ell}\Psi_1)_m}{\Psi_{1,m} z^m} &=
(e^{-\eta} -1)\frac{{\cal L}_{m,i}^{(\al)}
\tau_m}{\tau_m}\\
-\frac{((v_i^{(\al)})_u)^{\top}\Psi^*_2)_m}{\psi^*_{2,m}z^{-m}}&=
(e^{\tilde\eta} -1)
\frac{{\cal L}_{m,i}^{(\al)} \tau_{m}}{\tau_{m}} - \frac{{\cal
 L}_{m+1,i}^{(\al)} \tau_{m+1}}{\tau_{m+1}} + \frac{{\cal
 L}_{m,i}^{(\al)} \tau_m}{\tau_m} - (m+1) \delta_{i0}
\end{array}\leqno(5.7)
$$
\hfill for $i,m\in\BZ$,$\al=1,2$.

\medbreak

\noindent\underline{Proof of Lemma 5.2}: At first observe
that
$v_i^{(1)}$ and
$v_i^{(2)}$ defined in (5.1) equals, 
\begin{eqnarray*}
v_i^{(1)} &=& M_1L_1^{i+1} + (i+1)L_1^i +\sum_{r,s \geq 1} rc_{rs}
L_1^{i+r} L_2^s\\
v_i^{(2)} &=& M_2L_2^{i+1} -L_2^i -\sum_{r,s \geq 1} sc_{rs} L_1^r
L_2^{i+s}.
\end{eqnarray*}
Then using (1.27) and (1.27'), (4.13,(iii)), Corollary 1.2.1 and the
notations (5.3) and (5.4), we have the following:
\begin{eqnarray*}
\frac{{(\BY_{{v_i}^{(\al)}} \Psi_1)}_m}{\Psi_{1,m} z^m} 
&=& (-1)^{\al}\frac{{((v_i^{(\al)})_{\ell} \Psi_1)}_m}{\Psi_{1,m} z^m} =
(e^{-\eta}-1)\frac{{\cal K}_{m,i}^{(\al)} \tau_m}{\tau_m} 
 = (e^{-\eta}-1)\frac{{\cal L}_{m,i}^{(\al)} \tau_m}{\tau_m} \\
\frac{{(\BY_{{v_i}^{(\al)}} \Psi_2^{\ast})}_m}{\Psi_{2,m}^{\ast}
z^{-m}}&=&(-1)^{\al}\frac{{((v_i^{(\al)})_u)^{\top})\Psi_2^{\ast})}_m}
{\Psi_{2,m}^{\ast}
z^{-m}}\\
&=& (e^{\tilde\eta}-1)\frac{{\cal K}_{m,i}^{(\al)}
\tau_m}{\tau_m}
 -  \frac{{\cal K}_{m+1,i}^{(\al)} \tau_{m+1}}{\tau_{m+1}}
 + \frac{{\cal K}_{m,i}^{(\al)} \tau_m}{\tau_m}, \\
&=& (e^{\tilde \eta}-1)\frac{{\cal L}_{m,i}^{(\al)}
\tau_m}{\tau_m}
 -  \frac{{\cal L}_{m+1,i}^{(\al)} \tau_{m+1}}{\tau_{m+1}}
 + \frac{{\cal L}_{m,i}^{(\al)} \tau_m}{\tau_m} - (m+1)
\delta_{i0}, 
\end{eqnarray*}
where the term $(m+1)\delta_{i0}$ is caused by the constant which
distinguishes $\LR^{(\al)}_{m,i}$ from $\KR^{(\al)}_{m,i}$, ending
the proof of Lemma 5.2.

\medbreak

\medbreak\noindent\underline{Proof of Theorem 5.1}: That the matrices
$v_i^{(\al)}$ form an algebra 
$Vir^+\otimes Vir^+$ for $\{\,,\,\}$ with structure relations
(5.5) follows immediately from Theorem 2.3. Since the maps
$$
v,\{\,,\,\}\mapsto\BY_v,[\,,\,]
$$
and
$$
\BY_v\mbox{ acting on }L\mapsto\BY_v\mbox{ acting on }\tau
$$
are Lie algebra homomorphisms, the vector fields $\BY_{v_i^{(\al)}}$ induce on
$\tau$ an algebra $Vir^+\otimes Vir^+$ of constraints, as well.

\medbreak

For string-orthogonal polynomials, the matrices $L_1,M_1,L_2$ and
$M_2^*$ satisfy the string relations (4.7), implying, upon
multiplication by $L_1^{i+1}$ and $L_2^{i+1}$ respectively, that
both $v_i^{(1)}$ and $v_i^{(2}=0$ for $i\geq -1$; thus the right
hand side of (5.7) vanishes. But the terms on the right hand
side of (5.7), containing
$e^{-\eta}-1$ and
$e^{\tilde\eta}-1$, are Taylor series in $z^{-1}$ and $z$
respectively, without independent term, whereas the remaining
part is independent of $z$; therefore we have
$$\frac{\LR_{m,i}^{(\al)}\tau_m}{\tau_m}=a^{(\al)}(m,i,c),\quad\quad
m\geq 1\mbox{ and }\al =1,2, $$
is a function of $m,i,c=(c_{ij}$, independent of $t$ and
$s.$ Substituting the latter into the right hand side of (5.7),
yields the difference relation
$$
\frac{\LR_{m+1,i}^{(\al)}\tau_{m+1}}{\tau_{m+1}}-
\frac{\LR_{m,i}^{(\al)}\tau_m}{\tau_m}
+(m+1)\dt_{i0}=a^{(\al)}(m+1,i,c)-a^{(\al)}(m,i,c)+(m+1)
\dt_{i0}=0,\leqno{(5.8)}
$$
for $m\geq 0$, $i\geq -1$ and $\al=1,2$. 
Moreover, since $\LR_{0,i}^{(\al)}$ (for $i\geq -1$) is a differentiation
operator, and since $\tau_0=1$, we have the boundary condition
$$
a^{(\al)}(0,i,c)=\frac{\LR_{0,i}^{(\al)}\tau_0}{\tau_0}=0,
\quad\quad\mbox{for }i\geq -1.\leqno{(5.9)}
$$
So, the difference relation (5.8) together with the boundary
condition (5.9) implies
$$
a^{\al}(m,i,c)=-\frac{m(m+1)}{2}\dt_{i0},\quad\quad m\geq 0,i\geq
-1,
$$establishing Theorem 5.1.

\proclaim Corollary 5.1.1. For the two-matrix integral
$\tau_N$, relations (5.6) are equivalent to
$$
\left(J_i^{(2)}+(2N+i+1)J_i^{(1)}+N(N+1)J_i^{(0)}+2\sum_{r,s\geq
1}rc_{rs} \sum^{N-1}_{\al
=0}(L_1^{i+r}L_2^s)_{\al\al}\right)\tau_N=0
$$
$$
\left(\tilde J_i^{(2)}-(2N+i+1)\tilde
J_i^{(1)}+N(N+1)\tilde J_i^{(0)}+2\sum_{r,s\geq
1}sc_{rs}\sum^{N-1}_{\al
=0}(L_1^rL_2^{s+i})_{\al\al}\right)\tau_N=0,
$$ \hfill for $i\geq -1$ and $N\geq 0.$
\medbreak\noindent In particular, when $V_{12}=cyz$, we have
\begin{eqnarray*}
\left(J_i^{(2)}+(2N+i+1)J_i^{(1)}+N(N+1)J_i^{(0)}\right)\tau_N+
2c~p_{i+N}
(\tilde\pl_t)p_{N}(-\tilde\pl_s)\tau_1\circ\tau_{N-1}&=&0  \\
\\
\left(\tilde J_i^{(2)}-(2N+i+1)\tilde
J_i^{(1)}+N(N+1)\tilde J_i^{(0)}\right)\tau_N +2c~p_{N}
(\tilde\pl_t)p_{i+N}(-\tilde\pl_s)\tau_1\circ\tau_{N-1}&=&0
\end{eqnarray*} 
or, in terms of the moments $\mu_{ij}$,
\begin{eqnarray*}
\left(J_i^{(2)}+(2N+i+1)J_i^{(1)}+N(N+1)J_i^{(0)}\right)\tau_N+
2c\renewcommand{\arraystretch}{0.5}
\begin{array}[t]{c}
\sum\\
{\scriptstyle k+k'= N+i}\\
{\scriptstyle \ell+\ell'=N}\\
{\scriptstyle k,k',\ell,\ell'\geq 0}
\end{array}
\renewcommand{\arraystretch}{1}\mu_{k\ell}p_{k'}
(-\tilde\pl_t)p_{\ell'}(\tilde\pl_s)\tau_{N-1}
&=&0  \\
\\
\left(\tilde J_i^{(2)}-(2N+i+1)\tilde
J_i^{(1)}+N(N+1)\tilde J_i^{(0)}\right)\tau_N +2c\renewcommand{\arraystretch}{0.5}
\begin{array}[t]{c}
\sum\\
{\scriptstyle k+k'= N}\\
{\scriptstyle \ell + \ell'=N+i}\\
{\scriptstyle k,k',\ell, \ell'\geq 0}
\end{array}
\renewcommand{\arraystretch}{1}\mu_{k \ell}p_{k'}
(-\tilde\pl_t)p_{\ell'}(\tilde\pl_s)\tau_{N-1}
&=&0
\end{eqnarray*} 
\hfill for $i\geq -1$ and $N\geq 1.$

\medbreak\noindent\underline{Proof}: It suffices to replace the
partial derivatives\footnote{$\widehat{h_i}$ means with $h_i$
omitted}
\begin{eqnarray*}
\frac{\pl \tau_N}{\pl c_{\al\be}}&=&N!\sum_{i=0}^{N-1}\frac{\pl h_i}{\pl
c_{\al\be}}h_0\pp\widehat{h_i}\pp h_{N-1},\quad\mbox{using Theorem
3.2,}\\
&=&N!\sum_{i=0}^{N-1}(L_1^{\al}L_2^{\be})_{ii}h_ih_0\pp\widehat{h_i}\pp
h_{N-1}, \quad\mbox{using Proposition 1.1,}\\
&=&N!\sum_{i=0}^{N-1}(L_1^{\al}L_2^{\be})_{ii}\prod_0^{N-1}h_k\\
(*)\hspace{1cm}&=&\tau_N
\sum_{i=0}^{N-1}(L_1^{\al}L_2^{\be})_{ii},\hspace*{.5cm}\mbox{using (3.9),}\\
(**)\hspace{1cm}&=&p_{\al+N-1}(\tilde\pl_t)p_{\be
+N-1}(\tilde\pl_s)\tau_1\circ\tau_{N-1}, \hspace*{.5cm}\mbox{using theorem
4.3,}
\end{eqnarray*}
in the expression $\LR_{m,i}^{(\al)}$ (see (5.3) and (5.4)) of
(5.6) by the multiplication operator (*). For
$V_{12}=cyz$, one replaces $\frac{\pl \tau_N}{\pl c_{\al \be}}$ by the
Hirota-type expression (**) and then one sets all
$c_{ij}=0$, for all
$i,j\geq 1$, except
$c_{11}=c$; the second set of equations uses the formula in Remark 1,
following Theorem 4.3, thus ending the proof of  Corollary 5.1.1.

\bigbreak 

Note that, since
$$
(\frac{\pl}{\pl z})^pz^n=\sum_{k=0}^{p\wedge n}\MAT{1}p\\k\mat
(n)_k\, z^{n-k}(\frac{\pl}{\pl z})^{p-k}=: \sum_{k=0}^{p\wedge
n}\al_k^{(n,p)}z^{n-k}(\frac{\pl}{\pl z})^{p-k},\quad p,n\geq 0
$$ we have using (1.20), that
\begin{eqnarray*}
[L_1^n,M_1^p]&=&\sum_{k=1}^{p\wedge
n}\al_k^{(n,p)}M_1^{p-k}L_1^{n-k},\quad\mbox{ with
}\al_k^{(n,p)}:=\MAT{1}p\\k\mat(n)_k\\
(M_2-L_2^{-1})^n&=&\sum^n_{k=0}\be_k^{(n)}M_2^{n-k}L_2^{-k},
\quad\mbox{ with }\be_k^{(n)}:=\MAT{1}n\\k\mat 
(-1)^k,\quad\be_0^{(n)}=1.
\end{eqnarray*}
Defining 
$$
\bar\al_k^{(i,j)}:=\frac{\al_k^{(i,j)}}{(-c)^j(j-k+1)}\mbox{ and
}\bar\be_k^{(i)}=\frac{\be_k^{(i)}}{(i-k+1)c^i},\leqno{(5.10)} $$
we now state

\proclaim Theorem 5.2. For $V_{12}=cyz$, we have
$$
\Bigl(\sum_{k=0}^{i\wedge
j}\bar\al_k^{(i,j)}W_{m,i-j}^{(j-k+1)}+\sum_0^i\bar\be_k^{(i)}
\tilde W_{m-1,j-i}^{(i-k+1)}
\Bigr)\tau_m=\frac{i!}{(-c)^i}\dt_{ij}\tau_m,\quad\quad
i,j\geq 0 .$$ 

\medbreak\noindent\underline{Proof}: For $V_{12}=cyz$, the string
equations
$v_{-1}^{(\al)}=0$ (see (5.1)) reduce to
$$
M_1+cL_2=0\quad\mbox{ and }\quad M_2^{\ast
\top}+cL_1=0,\leqno{(5.11)} $$
implying
$$(-\frac{1}{c})^jL_1^iM_1^j-L_1^iL_2^j=0\quad\mbox{ and }\quad
(\frac{1}{c})^i(-M_2^{\ast \top})^iL_2^j-L_1^iL_2^j=0.
 $$
Subtracting we find, upon replacing $M_2^{\ast \top}$ by the
expression (5.2),
 \begin{eqnarray*}
0&=&(-\frac{1}{c})^jL_1^iM_1^j-(\frac{1}{c})^i(-M_2^{\ast\top})^i
L_2^j\\
&=&(-\frac{1}{c})^jL_1^iM_1^j-(\frac{1}{c})^i(M_2-
L_2^{-1})^iL_2^j\\
&=&\sum_{k=0}^{i\wedge
j}\frac{\al_k^{(i,j)}}{(-c)^j}M_1^{j-k}L_1^{i-k}-\sum_{k=0}^i
\frac{\be_k^{(i)}}{c^i}M_2^{i-k}L_2^{j-k}.
 \end{eqnarray*}
Applying Corollary 1.2.1, we have, using the notation
(5.10), 
$$
\begin{array}{ll}
0 &= -\frac{\Biggl(\Bigl(\sum_{k=0}^{i\wedge
j}\frac{\al_k^{(i,j)}}{(-c)^j}M_1^{j-k}L_1^{i-k}-\sum_{k=0}^i
\frac{\be_k^{(i)}}{c^i}M_2^{i-k}
L_2^{j-k}\Bigr)_{\ell}\Psi_1\Biggr)_m}
{\Psi_{1,m}z^m}\\[3mm]
&=(e^{-\eta}-1)\frac{\Bigl(\sum_0^{i\wedge
j}\bar\al_k^{(i,j)}W_{m,i-j}^{(j-k+1)}+\sum_0^i
\bar\be_k^{(i)}\tilde W_{m-1,j-i}^{(i-k+1)}\Bigr)\tau_m}{\tau_m}
\end{array}\leqno(5.12)
$$
 and
$$
\begin{array}{ll}
0&=\frac{-\Biggl(\Bigl(\Bigl(\sum_{k=0}^{i\wedge
j}\frac{\al_k^{(i,j)}}{(-c)^j}M_1^{j-k}L_1^{i-k}-\sum_{k=0}^i
\frac{\be_k^{(i)}}{c^i}M_2^{i-k}
L_2^{j-k}\Bigr)_u\Bigr)^{\top}\Psi_2^{\ast}\Biggr)_m}
{\Psi_{2,m}^{\ast}z^{-m}}\\[3mm]
&=e^{\tilde\eta}\frac{\Bigl(\sum_0^{i\wedge
j}\bar\al_k^{(i,j)}W_{m,i-j}^{(j-k+1)}+\sum_0^i
\bar\be_k^{(i)}\tilde
W_{m-1,j-i}^{(i-k+1)}\Bigr)\tau_m}{\tau_m}  \\[3mm] 
& \quad -\frac{\Bigl(\sum_{k=0}^{i\wedge
j}\bar\al_k^{(i,j)}W_{m+1,i-j}^{(j-k+1)}+\sum_{k=0}^i
\bar\be_k^{(i)}\tilde
W_{m,j-i}^{(i-k+1)}\Bigr)\tau_{m+1}}{\tau_{m+1}}. 
\end{array}\leqno(5.13)
$$
Upon using the same argument on (5.12) and (5.13) as in the proof
of Theorem 5.1, one shows the ratio is independent of $t$ 	and
$s$; so 
$$
\frac{\Bigl(\sum_0^{i\wedge
j}\bar\al_k^{(i,j)}W_{m,i-j}^{(j-k+1)}+\sum_0^i
\bar\be_k^{(i)}\tilde
W_{m-1,j-i}^{(i-k+1)}\Bigr)\tau_m}{\tau_m}=:a^{ij}
(m,c),\leqno{(5.14)}
$$
and according to (5.13),
$$
a^{ij}(m+1,c)=a^{ij}(m,c),\quad\quad\quad m=0,1,2,\pp
$$
Then, we compute (5.14) for $m=0$:
\medbreak
\noindent $a^{ij}(0,c)$
\begin{eqnarray*}
a^{ij}(0,c)&=&\frac{\Bigl(\sum_{k=0}^{i\wedge
j}\bar\al_k^{(ij)}W_{0,i-j}^{(j-k+1)}+\sum_{k=0}^i
\bar\be_k^i\tilde
W_{-1,j-i}^{(i-k+1)}\Bigr)\tau_0}{\tau_0}\\
&=&\frac{\Bigl(\sum_{k=0}^{i\wedge
j}\bar\al_k^{(ij)}(0)_{j-k+1}\delta_{i,j}+\sum^i_{k=0}
\bar\be_k^i(1)_{i-k+1}\delta_{i,j}\Bigr)\tau_0}{\tau_0}\\
& &\quad\quad +\frac{\Bigl(\mbox{(differential
operator)}+\mbox{(polynomial
in\,}t)\Bigr)\tau_0}{\tau_0},\quad\mbox{using (1.49),}\\
&=&\bar\be_i^i(1)_1\dt_{i,j}+\mbox{(polynomial in
}t)\quad\quad\mbox{since }\tau_0=1, (0)_{j-k+1}=0\\
&=&\frac{(i)_i(-1)^i}{c^i}\dt_{ij},\quad\quad\mbox{since
$a^{ij}(0,c)$ is independent of }t\\ &=&\frac{i!}{(-c)^i}\dt_{ij}.
\end{eqnarray*}

\end{document}